%% file: sample701.tex
\documentclass[trackchanges,twocolumn]{aastex701}

\usepackage{amsfonts, amsmath, amssymb, multirow}

\newcommand{\Sg}{Sgr~A*}
\shorttitle{MAXI J1744-294 Linear Polarization Detection and Evolution}
\shortauthors{Michail et al.}
\submitjournal{ApJ}

\begin{document}

\title{Detection and Evolution of Linear Polarization of the Galactic Center Transient MAXI J1744-294}

\author[orcid=0000-0003-3503-3446,sname=Michail, gname=Joseph]{Joseph M. Michail}
\altaffiliation{NSF Astronomy \& Astrophysics Postdoctoral Fellow}
\affiliation{Center for Astrophysics $|$ Harvard \& Smithsonian, 60 Garden Street, Cambridge, MA 02138, USA}
\email[show]{joseph.michail@cfa.harvard.edu}  

\author[orcid=0000-0002-9156-2249,sname='von Fellenberg', gname=Sebastiano]{Sebastiano D. von Fellenberg}
\altaffiliation{Feodor Lynen Fellow}
\affiliation{Canadian Institute for Theoretical Astrophysics, University of Toronto, 60 St.\ George Street, Toronto, ON M5S 3H8, Canada}
\affiliation{Max Planck Institute for Radioastronomy, auf dem H{\"u}gel 69, Bonn, Germany}
\email[show]{s.fellenberg@utoronto.ca}  

\author[0000-0001-9641-6550]{Mayura Balakrishnan}
\affiliation{Department of Physics, McGill University, 3600 Rue University, Montr\'{e}al, Qu\'{e}bec, H3A 2T8, Canada}%
\affiliation{Trottier Space Institute at McGill, 3550 Rue University, Montr\'{e}al, Qu\'{e}bec, H3A 2A7, Canada}
\email{mayura.balakrishnan@mcgill.ca}

\author[0000-0003-4056-9982]{Geoffrey C. Bower}
\affiliation{East Asian Observatory, 660 N. A'ohoku Pl., Hilo, HI, 96720, USA}
\affiliation{Academia Sinica Institute of Astronomy and Astrophysics, 645 N. A'ohoku Pl., Hilo, HI, 96720, USA}
\affiliation{Department of Physics and Astronomy, University of Hawaii at Manoa, 2505 Correa Road, Honolulu, HI, 96822, USA}
\email{gbower@asiaa.sinica.edu.tw}

\author[0000-0001-8921-3624]{Nicole M. Ford}
\affiliation{Department of Physics, McGill University, 3600 Rue University, Montr\'{e}al, Qu\'{e}bec, H3A 2T8, Canada}%
\affiliation{Trottier Space Institute at McGill, 3550 Rue University, Montr\'{e}al, Qu\'{e}bec, H3A 2A7, Canada}
\email{nicole.ford@mail.mcgill.ca}

\author[0009-0004-8539-3516]{Zach Sumners}
\affiliation{Department of Physics, McGill University, 3600 Rue University, Montr\'{e}al, Qu\'{e}bec, H3A 2T8, Canada}%
\affiliation{Trottier Space Institute at McGill, 3550 Rue University, Montr\'{e}al, Qu\'{e}bec, H3A 2A7, Canada}
\email{ronald.sumners@mail.mcgill.ca}

\author[0000-0002-0670-0708]{Giovanni G. Fazio}
\altaffiliation{Deceased after manuscript submission, February 2026}
\affiliation{Center for Astrophysics $|$ Harvard \& Smithsonian, 60 Garden Street, Cambridge, MA 02138, USA}
\email{gfazio@cfa.harvard.edu}

\author[0000-0001-6803-2138]{Daryl Haggard}
\affiliation{Department of Physics, McGill University, 3600 Rue University, Montr\'{e}al, Qu\'{e}bec, H3A 2T8, Canada}%
\affiliation{Trottier Space Institute at McGill, 3550 Rue University, Montr\'{e}al, Qu\'{e}bec, H3A 2A7, Canada}
\email{daryl.haggard@mcgill.ca}

\author[0000-0002-5599-4650]{Joseph L. Hora}
\affiliation{Center for Astrophysics $|$ Harvard \& Smithsonian, 60 Garden Street, Cambridge, MA 02138, USA}
\email{jhora@cfa.harvard.edu}

\author[orcid=0000-0002-3490-146X,sname=Keating, gname=Garrett]{Garrett K. Keating}
\affiliation{Center for Astrophysics $|$ Harvard \& Smithsonian, 60 Garden Street, Cambridge, MA 02138, USA}
\email{garrett.keating@cfa.harvard.edu}

\author[orcid=0000-0002-4090-8000]{J. D. Livingston}
\affiliation{Max Planck Institute for Radioastronomy, auf dem H{\"u}gel 69, Bonn, Germany}
\email{jack.david.livingston+academic@gmail.com}

\author[0000-0001-9564-0876]{Sera Markoff}
\affiliation{Anton Pannekoek Institute for Astronomy, University of Amsterdam, Science Park 904, 1098 XH Amsterdam, The Netherlands}
\affiliation{Gravitation and Astroparticle Physics Amsterdam Institute, University of Amsterdam, Science Park 904, 1098 XH 195 196 Amsterdam, The Netherlands}
\affiliation{Institute of Astronomy, University of Cambridge, Madingley Road, Cambridge CB3 0HA, United Kingdom}
\email{S.B.Markoff@uva.nl}


\author[0000-0002-7301-3908]{Bart Ripperda}
\affiliation{Canadian Institute for Theoretical Astrophysics, University of Toronto, 60 St.\ George Street, Toronto, ON M5S 3H8, Canada}
\affiliation{Dunlap Institute for Astronomy and Astrophysics, University of Toronto, 50 St.\ George Street, Toronto, ON M5S 3H4, Canada}
\affiliation{Department of Physics, University of Toronto, 60 St. George Street, Toronto, ON M5S 1A7, Canada.}
\email{bartripperda@gmail.com}

\author[0000-0002-0947-569X]{Sophia S\'anchez-Maes}
\affiliation{University of Maryland, College Park, MD 20742, USA}
\email{sophiasm@umd.edu}

\author[]{Howard A. Smith}
\affiliation{Center for Astrophysics $|$ Harvard \& Smithsonian, 60 Garden Street, Cambridge, MA 02138, USA}
\email{hsmith@cfa.harvard.edu}

\author[0000-0002-9895-5758]{S. P. Willner}
\affiliation{Center for Astrophysics $|$ Harvard \& Smithsonian, 60 Garden Street, Cambridge, MA 02138, USA}
\email{swillner@cfa.harvard.edu}

\author[orcid=0000-0002-1317-3328,sname=Zhao, gname=Jun-Hui]{Jun-Hui Zhao}
\affiliation{Center for Astrophysics $|$ Harvard \& Smithsonian, 60 Garden Street, Cambridge, MA 02138, USA}
\email{jzhao@cfa.harvard.edu}

\begin{abstract}
MAXI J1744$-$294, likely a low-mass X-ray binary system, is a Galactic-center transient source, detected at radio and X-ray wavelengths,  located approximately $19\arcsec$ southeast of \Sg. We report the first detection of its variable linear polarization in four epochs spanning 2025 Apr 04--09. The normalized 33 and 43~GHz Stokes parameters $q$ and $u$ over the four epochs imply a common Faraday rotation screen with a rotation measure RM $=-63\,606^{+844}_{-861}$ radians~m$^{-2}$, the third largest RM detected within the Galaxy. The RM is consistent with that of the Galactic center magnetar PSR J1745$-$2900, giving the first direct evidence that MAXI J1744 lies within the Galactic center region, is bound to \Sg, and therefore, is part of the nuclear star cluster. The uniformity in the Galactic center Faraday screen suggests that \Sg's $\approx-10^5$ rad m$^{-2}$ RM is intrinsic rather than originating from an unrelated line-of-sight source. On 2025 Apr 06, we detected a secondary polarized component with an additional RM $\approx-6000$ rad m$^{-2}$, which was not seen at any other epoch. Assuming this secondary component primarily cools by synchrotron radiation, the implied local magnetic field strength is $\sim$15--30 gauss. In the context of a jetted X-ray binary progenitor, the additional RM screen and magnetic field strength are explainable with a short-lived knot in a putative jet.
\end{abstract}


\section{Introduction}
MAXI J1744$-$294 (MAXI J1744) is the most recent Galactic center (GC) transient source. It was discovered in X-rays by the Monitor of All-sky X-ray Image (MAXI) on board the International Space Station in early 2025 January \citep{2025ATel_maxi_kudo}. Given the large positional uncertainty of the MAXI instrument, followup observations by the X-ray NinjaSat satellite were consistent with a new transient, rebrightening events from two X-ray sources near \Sg\ previously in outburst, or potentially \Sg\ itself \citep{2025ATel_maxi_kudo, 2025ATel_maxi_nakajima, 2025ATel_ninjasat_watanabe}. Subsequent SWIFT monitoring \citep{2025ATel_swift_Heinke}, with its higher angular resolution, revealed a separate source southeast of \Sg. Archival X-ray analysis later indicated that MAXI J1744 is associated with the previously known transient SWIFT J174540.2$-$290037, first discovered in outburst in 2016 May with SWIFT \citep{Degenaar2016a,Degenaar2016b,Mandel2025,Mandel2026_atel}.  \citet{Mori2019} presented a comprehensive analysis of its X-ray (Chandra and NuSTAR) properties during the 2016 outburst. No published radio limits exist during the 2016 May outburst, and the X-ray flux decayed by 2016 mid-July. By coincidence, the VLA Sky Survey \citep[VLASS;][]{Lacy2020}  made a pilot observation near Sgr~A* on 2016 Aug 13. Data retrieved from the VLASS archive show no discernible source at the SWIFT J174540.2$-$290037 location and place a $3\sigma$ upper limit of $\sim$1 mJy beam$^{-1}$ at 3~GHz on the source's radio emission on that date.

\citet{Mandel2025} provided a comprehensive multiwavelength study of the 2025 outburst, focusing on X-ray light curves and spectra. The study includes NuSTAR, XMM-Newton, and SWIFT spectral observations and modeling, which were best fit by a compound blackbody accretion disk and a power-law, signaling a low-mass X-ray binary (LMXB) source in a soft state. Additional X-ray follow up included XRISM \citep[][]{2025ATel_xrism_mandel}, Einstein Probe \citep[][]{2025ATel_einstein_wang}, Chandra \citep{2025ATel_chandra_mandel}, and NICER \citep[2025 March][]{2025ATel_nicer_jaisawal}. Furthermore, the IXPE X-ray polarimeter placed a 3$\sigma$ upper limit of 1.3\% on the linear polarization across the 2--8 keV band \citep{Marra2025_IXPE}. \citet{Mandel2025} did not detect a near-infrared counterpart in Keck continuum ($\lambda\approx2.12$ and $2.27~\mu$m) or in Brackett-$\gamma$ data.

The first 2025 detection of a radio counterpart of MAXI J1744 was on 2025 Jan 25 by MeerKAT \citep{2025ATel_meerkat_grollimund} at L-band ($\approx$1.28~GHz) as a flux excess compared to previous images of ${\approx}110\pm50$~mJy embedded in the extended GC emission.  \citet{2025ATel17174_vla_michail} detected MAXI J1744 at 33 and 43~GHz with the National Radio Astronomy Observatory's Karl G. Jansky Very Large Array (VLA) on 2025 Apr 04, finding that the source's total intensity varied by a factor of $\approx$2--3 over subsequent observations on 2025 Apr 06, 07, and 09. Using full-track VLA observations of Sgr A* in 2024 April, \citet{2025ATel17174_vla_michail} placed $3\sigma\approx4$ mJy beam$^{-1}$ flux limits at 22, 33, and 44 GHz on MAXI J1744's quiescent radio flux prior to the 2025 January outburst. The 2025 VLA observations also showed that the radio spectral index changed significantly over these few days. This change was accompanied by increased X-ray hardness and hard X-ray (7.2--50~keV) count rate as probed by NuSTAR \citep{Mandel2025}.

This paper presents the first detection of MAXI J1744's radio polarization and rotation measure (RM) with the VLA at 33 and 43~GHz in 2025 April. These data were taken as part of our ongoing multiwavelength monitoring of the variability of \Sg\ and its surroundings \citep{vonFellenberg2025, Michail2025, Roychowdhury2025}. Section~\ref{sec:data} describes the calibration and imaging of the radio data. Section \ref{sec:rm_pl} analyzes the polarization properties and evolution and models the normalized Stokes $q$ and $u$ spectra to reveal the third-largest RM from a Galactic source.  Section~\ref{sec:disc} discusses the results, and Section \ref{sec:conclusion} summarizes our work and presents our conclusions.

\section{VLA Data and Calibration}\label{sec:data}

    \begin{figure*}[ht]
        \centering
        \includegraphics[width=1\linewidth]{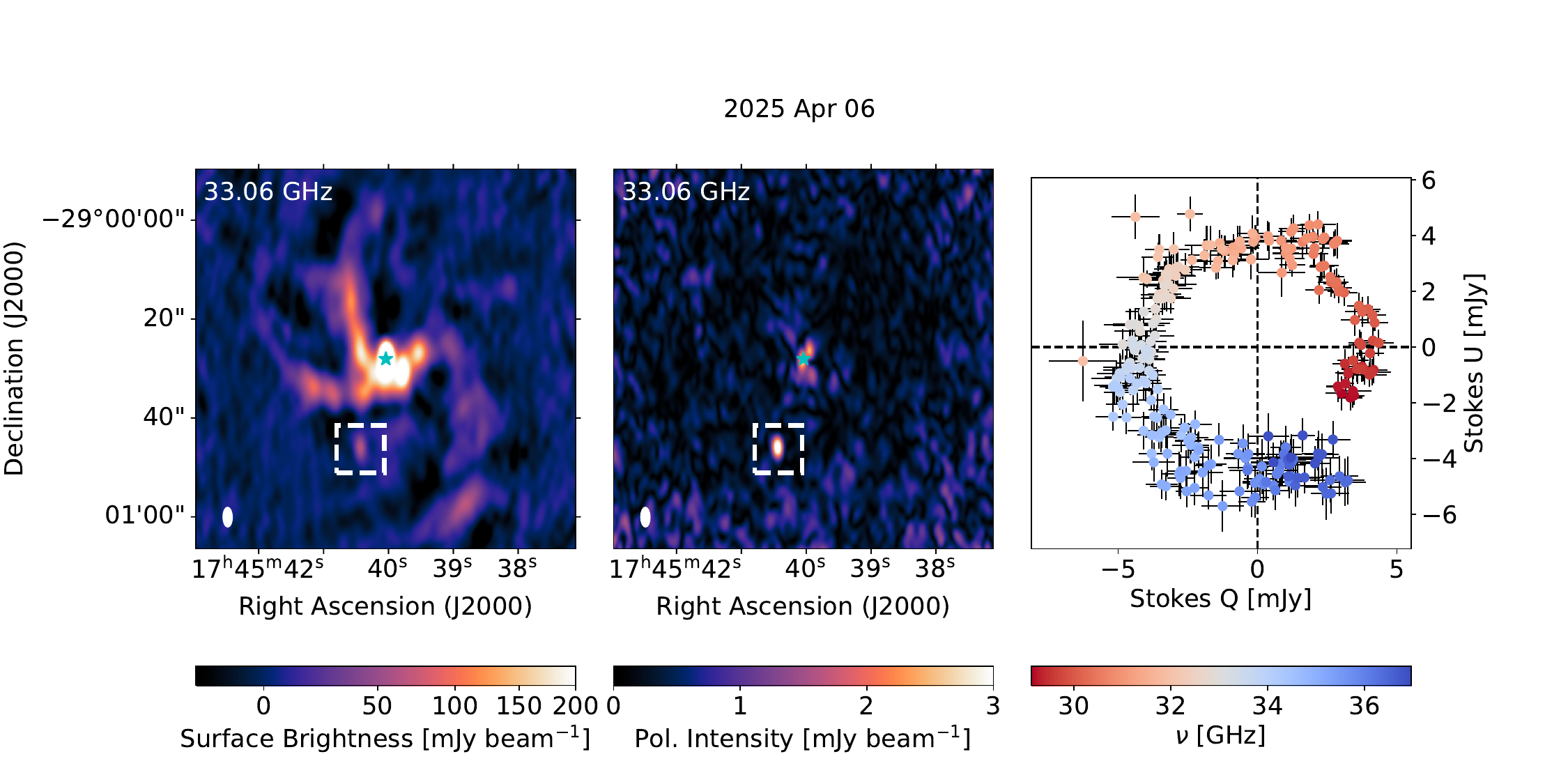}
        \caption{33 GHz images of \Sg\ and MAXI J1744 on 2025 Apr 06.  \textit{Left}: Stokes $I$ image. \Sg\ is marked with a cyan star, and MAXI J1744 is marked with a white dashed box. The synthesized beam is shown in the lower left corner. \textit{Center}: linearly polarized flux ($PI$) image. The markings are as in the left panel. \textit{Right}: Stokes $U$ vs.\ $Q$ of MAXI J1744. The colored points correspond to frequency as indicated in the color bar. The clockwise orientation of loop with increasing frequency corresponds to a negative rotation measure.}
        \label{fig:6apr_plots}
    \end{figure*}
    
    \subsection{Observing Setup and Calibration}
        \begin{deluxetable}{cccl}[t]
            \tablewidth{\linewidth}
            \tablenum{1}
            \tablecaption{VLA observing parameters for the data used in this analysis.\label{tab:obs}}
            \tablehead{\colhead{Date} & \colhead{On-source Time} & \colhead{Band}  & \colhead{Synthesized Beam} \\                 \colhead{}   &        \colhead{[hr]}      &       \colhead{}            &   \colhead{}}
            \startdata
                2025 Apr 04 & 3.39 & Q & $2\farcs57\times1\farcs40~\left(-4\fdg33\right)$\\
                2025 Apr 06 & 3.75 & Ka & $3\farcs93\times1\farcs81~\left(~~0\fdg80\right)$\\
                2025 Apr 07 & 3.39 & Q  & $3\farcs19\times1\farcs59~\left(-3\fdg02\right)$\\
                2025 Apr 09 & 3.75 & Ka  & $3\farcs82\times1\farcs83~\left(-1\fdg26\right)$\\
            \enddata
        \end{deluxetable}
        
        We obtained four full-track D-configuration observations of \Sg\ and its environs during 2025 April (project ID 25A-314, PI: Michail) as part of a multiwavelength campaign targeting \Sg\ and spanning radio through X-ray frequencies. Table~\ref{tab:obs} describes the observational parameters for the datasets used in this analysis. The observations used the 3-bit setup, providing full polarimetric products over 63 spectral windows (SPWs), each SPW composed of 64 2-MHz-wide channels. This gave 8~GHz of contiguous bandwidth between 29.1 and 37.0~GHz  (Ka band, 33~GHz) or 40.1 to 48.0~GHz  (Q band, 43~GHz).

        Standard high-frequency observational setups were completed, including regular pointing calibration required above 15~GHz. The Stokes $I$ data were processed through the default VLA pipeline in CASA 6.4.1 \citep{CASA2022}, and further manual processing to calibrate the polarization products followed the standard VLA polarization recipe.\footnote{\url{https://casaguides.nrao.edu/index.php/CASA_Guides:Polarization_Calibration_based_on_CASA_pipeline_standard_reduction:_The_radio_galaxy_3C75-CASA6.5.4}} We used J1331+3030 as the flux and absolute polarization-angle calibrator, J1733$-$1304 as the bandpass and instrumental polarization calibrator, and J1744$-$3116 as the complex gain calibrator. After calibration, data on \Sg\ underwent manual data flagging and two rounds of phase self-calibration with 3-second solution intervals. Phase self-calibration on baselines ${\geq} 30~k\lambda$ assumed a point source at the phase center. The data were imaged on all baselines, and the model components were used for phase self-calibration. Before imaging, we re-computed the statistical weights on the visibility data with \texttt{statwt}. Appendix~\ref{appx:pol_conventions} defines the polarization conventions used in this paper, and Appendix~\ref{appx:dterm} gives estimates of the residual instrumental polarization.

    \subsection{Imaging and Aperture Photometry}\label{ssec:imaging}
        Two \texttt{TCLEAN} imaging runs were needed to create observation-averaged full-polarization spectral cubes  of the region surrounding \Sg.  The first run imaged the Stokes $I$ and $V$ planes (\texttt{specmode=`cube'}, \texttt{stokes=`IV'}) at 32 MHz spectral resolution (\texttt{width=`32MHz'}). The \texttt{auto-multithresh} automatic masking algorithm \citep{automultithresh}, was required because of the extended Mini-spiral emission in Stokes $I$\null. Stokes $V$ was also imaged as a diagnostic of potential calibration- or antenna-based corruptions (such as beam squint). The second run imaged the Stokes $Q$ and $U$ planes (\texttt{stokes=`QU'}) with a manual mask composed of two circular,  $5\arcsec$-radius masks centered on \Sg\ and MAXI J1744.

        The $IQUV$ images for each observation epoch used a common restoring beam (\texttt{restoringbeam=`common'}) as listed in Table~\ref{tab:obs}. The cubes were non-interactively CLEANed with the multiscale deconvolver, \texttt{scales=[0, 5, 15] pix}, and a $3\sigma$ RMS stopping criterion. Additional parameters were Briggs weighting with \texttt{robust=0.5}, imaging limited to projected baselines ${\geq}10~\rm{k}\lambda$ to suppress emission with angular sizes ${\geq}21\arcsec$, and per-plane primary-beam correction (\texttt{pbcor=True}). The final 43~GHz data used a pixel size of 0\farcs24 and image size of 432 pixels (1\farcm73), and the 33~GHz data used 0\farcs32 pixels with image size 450 pixels (2\farcm40). The primary-beam-corrected cubes were exported from CASA format into FITS cubes for further analysis. The left and central panels of Figure~\ref{fig:6apr_plots} show single-epoch 33~GHz images of \Sg\ in Stokes $I$ (total intensity) and polarized intensity ($PI$).

        Stokes $I$, $Q$, and $U$ values for \Sg\ and MAXI J1744 were measured from aperture photometry with the \texttt{photutils} package \citep{astropy:2013, astropy:2018, astropy:2022}.  The chosen aperture was elliptical with the same position angle as the synthesized beam and 1.25$\times$ the radii (Table~\ref{tab:obs}).  The aperture centers were determined from the $PI$ image (e.g., the central panel of Figure~\ref{fig:6apr_plots}) at each epoch. $PI$ is preferred over the individual Stokes parameters because it has no large-scale component (unlike Stokes $I$), and $PI$ is positive definite, which is not necessarily true for Stokes $Q$ and $U$. Both properties make the centroiding procedure more accurate. The single aperture location per observation assumes that the total and polarized intensities are co-located (i.e., no core shift nor extended component), which is reasonable given the large restoring beams. On 2025 Apr 06, the Stokes $I$ spectrum of MAXI J1744 was dimmer by ${\sim}10$~mJy in a single baseband covering 31--33~GHz. This shift also appeared in \Sg's spectrum and in the Stokes $I$ spectra of J1733$-$1304 and J1744$-$3116, consistent with a flux bootstrapping calibration issue. This issue was limited to Stokes $I$ with no similar issues being detected in MAXI J1744's Stokes $Q$ and $U$ spectra. Moreover, neither the instrumental polarization solutions nor Stokes $Q$ and $U$ spectra for J1733$-$1304 or J1744$-$3116, produced using the same methods, showed any systematic variations in the 31--33~GHz linear polarization on 2025 Apr 06. Appendix~\ref{appx:fluxcorr} describes how the $I$ flux scaling for 2025 Apr 06 was corrected. The noise estimates on the measured Stokes parameters were calculated using the standard deviation in a nearby region which does not contain any significant sources or emission. 
        
        This paper's key results are based on normalized Stokes parameters (Stokes $q \equiv Q / I$ and $u \equiv U/ I$), which remove the Stokes $I$ spectral dependence from Stokes $Q$ and $U$ and may otherwise cause artifacts in the derived polarization properties if not accounted for \citep{Brentjens2005}.

\section{Rotation Measure and Polarization Properties}\label{sec:rm_pl}

    \begin{figure*}
        \centering
        \includegraphics[trim = 2.25cm 0 3.75cm 0, clip, width=0.85\linewidth]{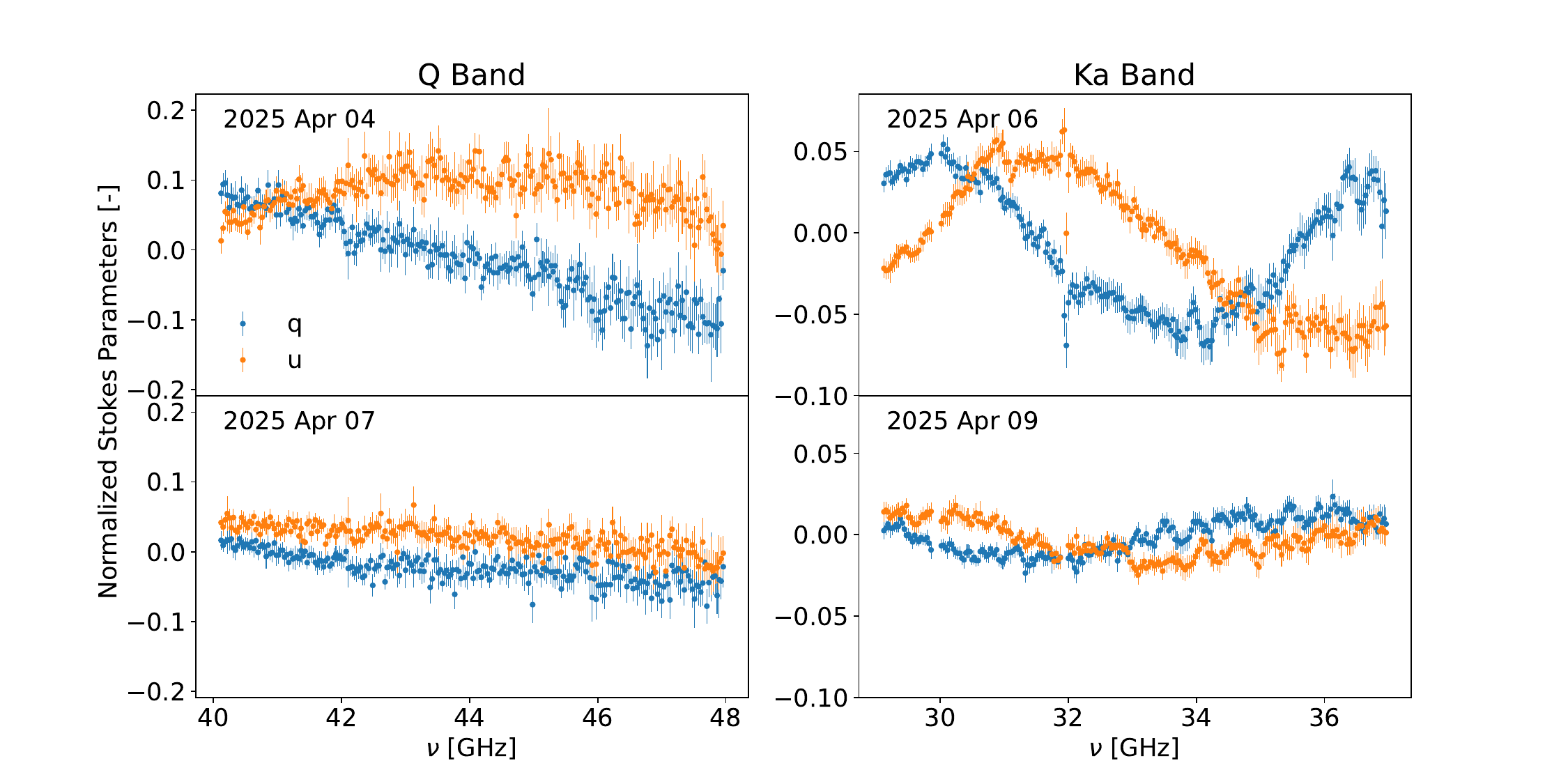}
        \caption{Normalized polarization parameters $q$ (blue) and $u$ (orange) of MAXI J1744 for each observation at 32 MHz spectral resolution. Y-axis ranges are shared between each frequency band.}
        \label{fig:qu_spectrum}
    \end{figure*}

    \subsection{Total Intensity and Polarization Variability}
    
        \begin{deluxetable}{lcccc}
            \tabletypesize{\scriptsize}
            \tablewidth{0pt} 
            \tablenum{2}
            \tablecaption{Total intensity and band-averaged linear polarization intensity and percent of MAXI J1744.\label{tab:intensities}}
            \tablehead{\colhead{Date} & \colhead{Stokes $I$\tablenotemark{a}} & \colhead{$\alpha$} & \colhead{$PI$} & \colhead{Pol. Percent} \\                 
                         \colhead{}   &  \colhead{[mJy]}   &      \colhead{} &   \colhead{[mJy]}  &      \colhead{[\%]}}
            \startdata
                \hline
                \multicolumn{5}{c}{33~GHz}\\
                \hline
                2025 Apr 06 & $74.9\pm0.6$ & $-1.58\pm0.12$ & $4.03\pm0.03$ & $5.45 \pm 0.07$ \\
                2025 Apr 09 & $94.6\pm0.4$& $-0.74\pm0.06$ & $1.27\pm0.03$ & $1.34 \pm 0.03$\\ 
                \hline
                \multicolumn{5}{c}{43~GHz}\\
                \hline
                2025 Apr 04 & $53.4\pm0.4$ & $~~0.08\pm0.13$  & $5.47\pm0.07$ & $10.12\pm0.13$ \\
                2025 Apr 07 & $73.9\pm0.3$ & $~~1.04\pm0.09$ & $2.74\pm0.06$ & $3.59 \pm 0.08$ \\
            \enddata
            \tablenotetext{a}{Listed flux densities are evaluated at $33$ and $43$~GHz for the Ka and Q bands, respectively. The in-band spectral index $\alpha$ is defined by $F_\nu\propto\nu^\alpha$.}
        \end{deluxetable}

        The right panel of Figure~\ref{fig:6apr_plots} shows the Stokes $U$ vs.\ $Q$ plot for the Ka-band observation on 2025 Apr 06; the loop is counterclockwise oriented with increasing frequency, exhibiting the presence of an intrinsic, negative RM across the band.
        Figure~\ref{fig:qu_spectrum} shows the normalized Stokes parameter spectra for each epoch across the full $\sim8$~GHz bandwidth. There are correlated spectral changes in the normalized Stokes parameters between each epoch, consistent with variable polarization properties on a $\sim$few-day cadence. To quantify the variability of the linear polarization, we calculated the band-averaged debiased $PI$ and polarization percents for each observation, which are listed in Table~\ref{tab:intensities}. (These quantities are defined in Appendix~\ref{appx:pol_conventions}\null.) The linear polarization fractions of MAXI J1744 on 2025 Apr 07 and 09 were approximately $3$ times lower than on 2025 Apr 04 and 06, whereas the $PI$ decreased only by a factor of $\sim$2 over the same period. The larger change in the polarization percent was exacerbated by the increasing flux density of $\sim40\%$ at 43~GHz and $\sim25\%$ at 33~GHz.\footnote{There is a typo in the 33~GHz flux density listed by \citet{2025ATel17174_vla_michail} on 2025 Apr 06, which initially showed a more drastic increase in the 33~GHz band flux density.} 
        
        Changes in the flux density were also commensurate with increases of the in-band spectral index ($\alpha$: $F_{\nu}\propto\nu^{\alpha}$),  calculated from the Stokes $I$ spectra (Section \ref{sec:data}) rebinned to 256 MHz resolution. We excluded $\nu\geq47.5$~GHz from the fitting on the 4th and 7th, as residual spectral roll-off is present in the spectra of MAXI J1744 and \Sg. There was a statistically significant hardening, i.e., an increase in $\alpha$, in the radio spectra across the 33~GHz ($\Delta\alpha_{33\rm{~GHz}} = 0.84\pm0.13$) and 43~GHz bands ($\Delta\alpha_{43\rm{~GHz}} = 0.96\pm0.16$). The changes in the spectral index across these two bands are statistically consistent, suggesting a single event or component was responsible the change in Stokes I.

        \begin{figure*}[t]
                \centering
                \includegraphics[width=0.85\linewidth]{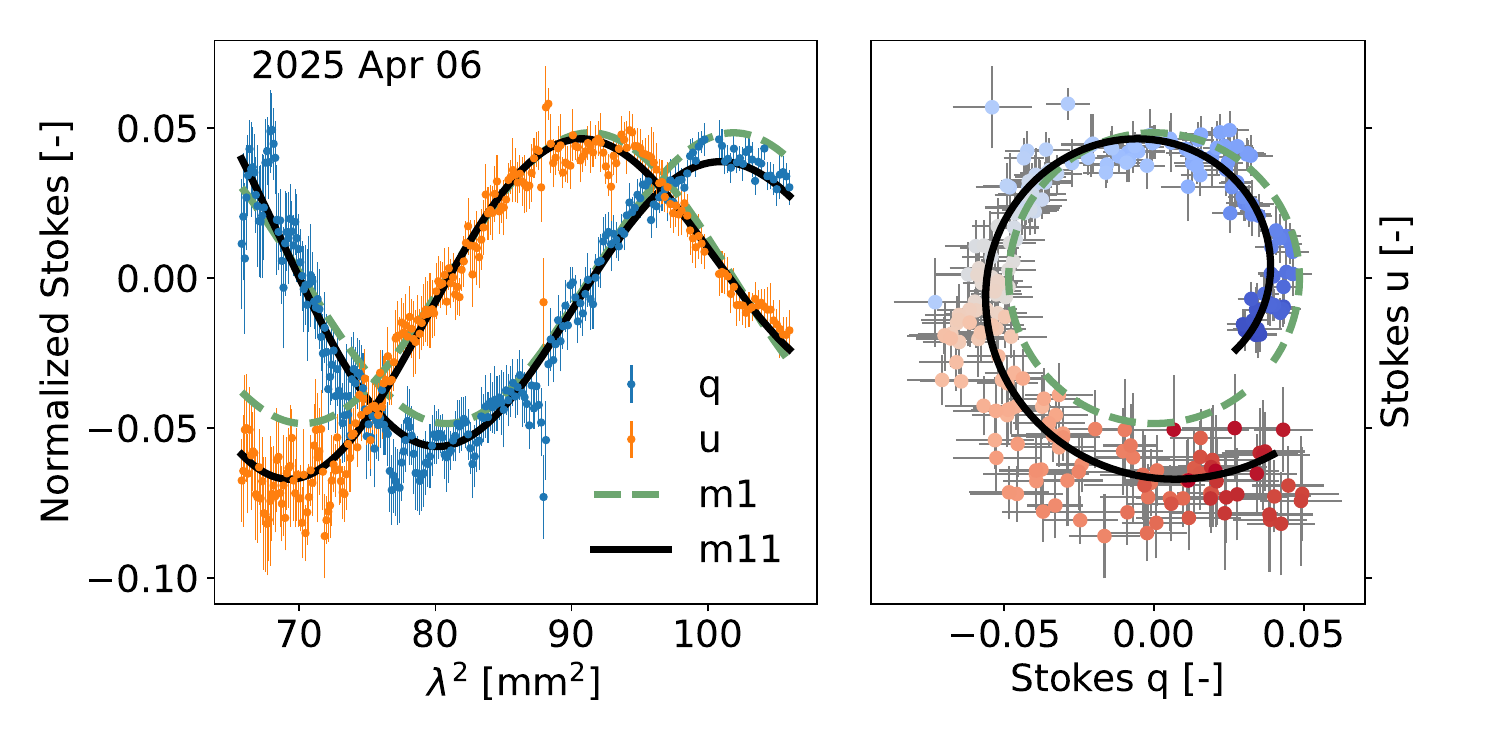}
                \caption{Normalized Stokes $q$ and $u$ parameters of MAXI J1744 on the 2025 Apr 06 epoch compared with the best-fit \texttt{m1} (dashed light green lines; single component) and \texttt{m11} (solid black lines; double component) models.  Appendix \ref{appx:results} gives more details about the models. \textit{Left}: Stokes $q$ and $u$ spectra. \textit{Right}: Stokes $q$ versus $u$ plot. The data-point colors correspond to the observing frequency, matching the color scheme in Figure~\ref{fig:6apr_plots}.}
                \label{fig:apr6_spec}
        \end{figure*}
        
    \subsection{Stokes q and u Spectral Fitting}
        \subsubsection{Per-Epoch Modeling}
            We utilized the \texttt{RM-Tools} \citep{RMTools2020}\footnote{\url{https://github.com/CIRADA-Tools/RM-Tools}} package to directly fit the Stokes $q$ and $u$ spectra for each observation. In contrast to fitting only the changes in the electric vector position angle (EVPA), this removes $n\pi$ ambiguities. Details about \texttt{RM-Tools} and the models used are in Appendix~\ref{appx:results}\null. In short, we fit each epoch with two models: 1) a single polarized component with one external Faraday screen, and 2) two polarized components (within the same beam) with two separate Faraday screens. Table~\ref{tab:qumodels} lists the best-fit parameters and goodness of fit quantities for each epoch and model. The polarization properties of MAXI J1744 are best represented by a single component on 2025 Apr 04, 07, and 09. However, two separate polarized components are significantly favored on 2025 Apr 06. The two-component model on Apr 06 is still preferred even when excluding the 31--33~GHz data because of the flux-scaling issue (Section~\ref{ssec:imaging}, Appendix~\ref{appx:fluxcorr}).  This test eliminates the possibility of a data-quality or systematic issue explaining the second component. Figure~\ref{fig:apr6_spec} shows the single- and double-component Stokes $q$ and $u$ model fits on this date, clearly demonstrating the need for a secondary component.
            
            The \texttt{RM-Tools} fits show a persistent Faraday screen with RM $\approx-63\,000$ rad m$^{-2}$, consistent with the intervening GC screen as probed by the GC magnetar PSR J1745$-$2900 \citep[e.g.,][]{Eatough2013,Desvignes2018}, yielding the first direct evidence that MAXI J1744 lies behind the GC screen and is most likely bound to \Sg. Additionally, the 2025 Apr 06 second component shows that the more highly polarized component has a higher Faraday depth, suggestive of a Faraday screen local to MAXI J1744's environment.

        \subsubsection{Joint-Epoch Modeling}
            \begin{figure}[h]
                \centering
                \includegraphics[width=\linewidth]{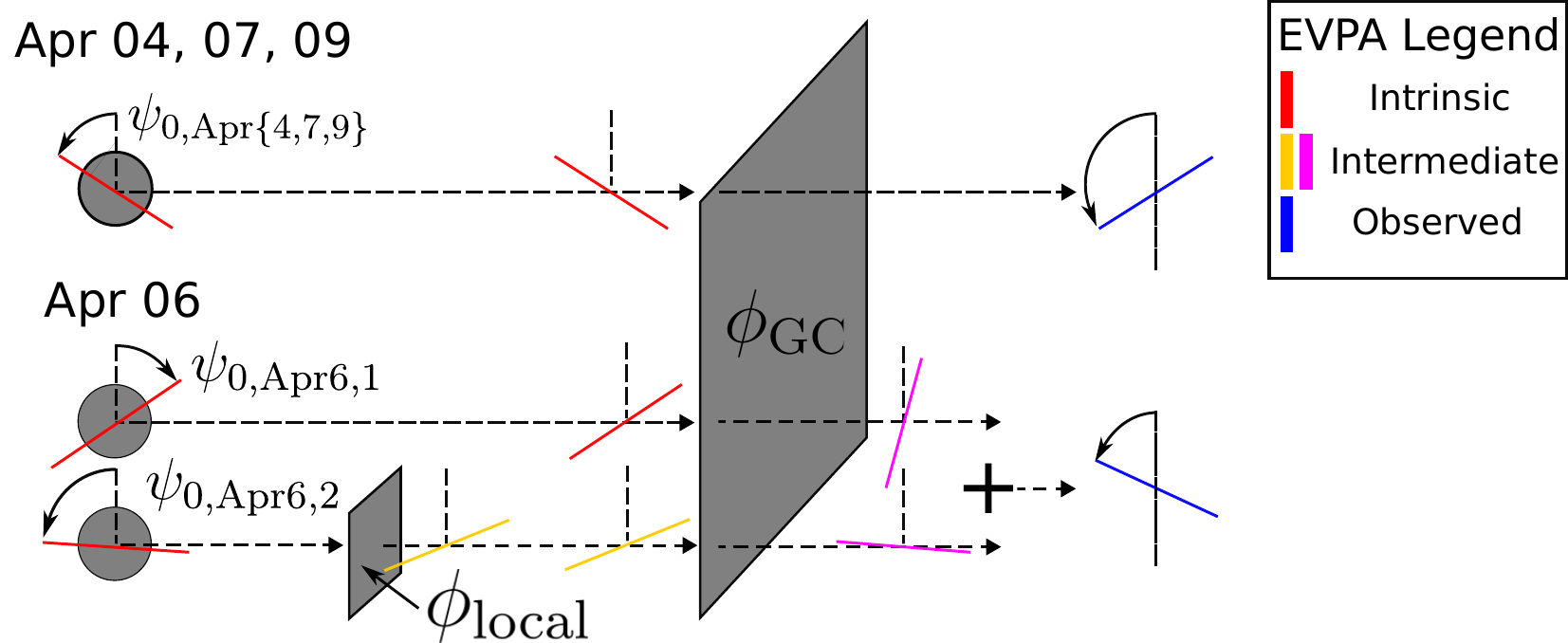}
                \caption{Schematic setup of the joint polarization fit among all four VLA epochs. The first, third, and final epochs are affected only by the single Galactic-center Faraday screen ($\phi_{\rm{GC}}$) whereas one of the two components in the second epoch has an additional secondary Faraday screen ($\phi_{\rm{local}}$). Intrinsic polarization angles for each day and component are denoted in red, while the measured polarization angles are in blue. Intermediate (unobserved) polarization angles are shown in yellow and magenta. ``+'' denotes that the sum of the two components gave the observed linear polarization on 2025 Apr 06. Physical scales are not accurately represented.}
                \label{fig:model_schematic}
            \end{figure}

            \begin{deluxetable}{lcc}
            \tabletypesize{\footnotesize}
            \tablewidth{\linewidth} 
            \tablenum{3}
            \tablecaption{Joint model best-fit parameters.\label{tab:jointmodel}}
            \tablehead{\colhead{Parameter} & \colhead{Prior} & \colhead{Posterior (68\% CI)}}
            \startdata
                $\phi_{\rm{GC}}$ [rad m$^{-2}$]          & $\mathcal{U}[-110\,000, 0]\tablenotemark{a}$ & $-63\,606^{+844}_{-861}$ \\
                $p_{0,\rm{Apr~4}}~[\%]$            & $\mathcal{U}[0, 100]$ & $9.84^{+0.16}_{-0.16}$\\
                $p_{0,\rm{Apr~6}, 1}~[\%]$         & $\mathcal{U}[0, 100]$ & $8.12^{+5.50}_{-2.82}$ \\
                $p_{0,\rm{Apr~7}}~[\%]$            & $\mathcal{U}[0, 100]$ & $3.63^{+0.11}_{-0.11}$   \\
                $p_{0,\rm{Apr~9}}~[\%]$            & $\mathcal{U}[0, 100]$ & $1.54^{+0.04}_{-0.04}$ \\
                $\psi_{0,\rm{Apr~4}}~[^\circ]$     & $\mathcal{U}[0, 180]\tablenotemark{b}$ & $39.75^{+2.44}_{-2.38}$  \\
                $\psi_{0, \rm{Apr~6}, 1}~[^\circ]$ & $\mathcal{U}[0, 180]$ & $79.95^{+6.11}_{-7.69}$  \\
                $\psi_{0,\rm{Apr~7}}~[^\circ]$     & $\mathcal{U}[0, 180]$ & $61.94^{+2.54}_{-2.56}$  \\
                $\psi_{0,\rm{Apr~9}}~[^\circ]$     & $\mathcal{U}[0, 180]$ & $62.13^{+4.32}_{-4.25}$  \\
                $\phi_{\rm{local}}$ [rad m$^{-2}$]          & $\mathcal{U}[-110\,000, 0]$ & $-6224^{+2119}_{-2482}$ \\
                $p_{0,\rm{Apr~6}, 2}~[\%]$         & $\mathcal{U}[0, 100]$ & $11.42^{+5.67}_{-3.19}$  \\
                $\psi_{0, \rm{Apr~6}, 2}~[^\circ]$ & $\mathcal{U}[0, 180]$ & $33.15^{+11.99}_{-10.37}$  \\
            \enddata
            \tablenotetext{a}{$\mathcal{U}[a,b]$ is a uniform prior on the inclusive interval $[a,b]$}
            \tablenotetext{b}{Priors on $\psi_0$ are periodic such that $\psi_0=180^\circ+n^\circ = n^\circ$.}
            \tablecomments{The goodness of fit measures for this fit are $\chi^2/\rm{DOF}=1161.35/1933$, BIC $=-10\,583.26$, and ln Evidence $=5277.68\pm0.33$ }
            \end{deluxetable}
            
            The consistent value of the GC-associated Faraday screen in the per-epoch fitting motivates a joint fit to all four epochs with a more sophisticated model. A schematic diagram of this setup is in shown in Figure~\ref{fig:model_schematic}. Using the results from the individual epoch fitting, we assume MAXI J1744 is best described on 2025 Apr 04, 07, and 09 by a single component with fixed $phi_{\rm GC}$ and other parameters varying by date:
                \begin{equation}
                    p_{\rm{Apr \{4,7,9\}}}\left(\lambda\right) = p_{0,\rm{Apr \{4,7,9\}}} e^{2i\left(\psi_{0,\rm{Apr \{4,7,9\}}} + \phi_{\rm{GC}}\lambda^2\right)},
                    \label{eq:general}
                \end{equation}
            where $p_{0,\rm{Apr \{4,7,9\}}}$ and $\psi_{0,\rm{Apr \{4,7,9\}}}$ are the intrinsic polarization fractions and EVPA on each of the days, respectively, and $\phi_{\rm{GC}}$ is the Faraday depth of the invariant GC screen.
            
            On 2025 Apr 06, we account for a secondary component with an additional, local Faraday screen ($\phi_{\rm{local}}$), which also passes through the GC screen:
                \begin{equation}
                    \begin{split}
                        p_{\rm{Apr~6}}\left(\lambda\right) = &p_{0,\rm{Apr~6},1} e^{2i\left(\psi_{0,\rm{Apr 6},1} +\phi_{\rm{GC}}\lambda^2\right)} +\\ &p_{0,\rm{Apr~6},2} e^{2i\left(\psi_{0,\rm{Apr 6},2} + \left(\phi_{\rm{GC}}+\phi_{\rm{local}}\right)\lambda^2\right)}.
                        \label{eq:apr6}
                    \end{split}
                \end{equation}
            Here $p_{0,\rm{{Apr~6}}}$ and $\psi_{0,\rm{Apr 6}}$ have the same meaning as in Equation~\ref{eq:general}, only for 2025 Apr 06. The ``1'' and ``2'' subscripts denote components that are affected only by the GC Faraday screen and by both the local and GC Faraday screens, respectively. $\phi_{\rm{GC}}$ is common to all observations. The joint fit had to be completed outside of \texttt{RM-Tools}, which is currently unable to do joint-epoch fitting. However, our implementation used the same underlying fitting code \citep[\texttt{dynesty};][]{dynesty} and cost function as \texttt{RM-Tools} for consistency. The best-fit joint-modeling parameters are shown in Table~\ref{tab:jointmodel}. 
            
            While our VLA observations detected the secondary polarized component on 2025 Apr 06, the observing frequency was too high to resolve any polarized substructures in the localized RM screen. Such polarized substructure is normally detected using RM synthesis \citep[e.g.,][]{OSullivan2012}. However, \citet{Dickey2019} showed the full-width half-maximum (FWHM) of the recovered RM synthesis spectrum is $\mathrm{FWHM} = 3.8/(\lambda^2_{\rm{max}}-\lambda_{\rm{min}}^2)$. For the 33~GHz VLA data, $\rm{FWHM} \approx 100\,000$ rad m$^{-2}\gg \left|\phi_2\right|$, rendering this impossible with the current data. Continued monitoring of MAXI J1744's polarized properties should be concentrated at lower frequencies, where polarized substructures can be resolved in the RM synthesis spectrum. However, such lower-frequency observations will contend with bandwidth depolarization from the GC Faraday screen, thus requiring high frequency resolution \citep[e.g.,][]{Bower1999}.
        
\section{Discussion}\label{sec:disc}
    \subsection{Temporal Evolution of the Primary Polarized Component}
        Table~\ref{tab:jointmodel} lists the joint-fit parameters for the ``primary'' polarization component (i.e., the component affected only by the GC Faraday screen) and the ``secondary'' component (i.e., the component with an additional, local Faraday screen). 
        The primary component's intrinsic linear polarization fraction exhibited a steady decrease over the four epochs, while the intrinsic EVPA changes were more complicated. Between 2025 Apr 04 and Apr 06, the intrinsic EVPA swung by $\approx$40$^\circ$ before shifting back by $\approx$10$^\circ$ between Apr 06 and Apr 07. It then stabilized through Apr 09, despite being measured at different frequencies. The trends in the polarization  coincide with trends in the spectral index (Table \ref{tab:intensities}): the intrinsic linear polarization decreased as the spectral index hardened and the flux density increased.

        In jetted sources, spectral hardening is typically interpreted as an increase in the optical depth of the jet, which can lower both the observed flux density \citep{Blandford1979} and the  linear polarization \citep{Longair2011}. The measured anti-correlation in the radio spectral index and intrinsic polarization is consistent with an increase in the jet's optical depth. However, the jet did not enter a fully optically thick regime, as $\alpha \not\approx 2.5$ \citep{ERA2016}, nor did the intrinsic EVPA exhibit the expected ${\sim}90^\circ$ rotation \citep{Longair2011}. The flux density did not follow the expected trend, as it increased despite the hardened radio spectrum. Observations of other jetted sources at $\sim$daily cadence \citep[e.g.,][]{Hannikainen2000} typically show an anti-correlation in the spectral index and flux density during individual flaring events, consistent with an increasing optical depth. However, \citet{Brocksopp2013} detected one instance where a hardened radio spectrum corresponded to an increased flux density and linear polarization at 5.5 and 9 GHz, which they argued can be explained by a forming compact jet. The overall evolution of MAXI J1744's radio polarization over these four epochs appears consistent with the formation and/or evolution of a compact jet and is also compatible with the lack of detectable X-ray polarization by IXPE, which observed between 2025 Apr 05 and Apr 08 \citep{Marra2025_IXPE}, although in other stellar-mass black hole binaries, the X-ray polarization may instead trace the accretion disk and/or corona rather than the jet \citep{Rodriguez2024}. Yet, $\sim$minute-timescale radio monitoring of V404 Cyg \citep{Hughes2023} found uncorrelated variability in the flux density and linear polarization evolution over $\sim$4 hours. Their results suggest changes in the flux density and polarization may be caused by a more complicated, multi-component emission model rather than a single emitting zone. The data presented here also suggest MAXI J1744 contains additional components (beyond the second polarized component on 2025 Apr 06) that are spatially-, spectrally-, and polarimetrically-unresolved by our observations. Moreover, $\sim$daily X-ray monitoring with NuSTAR, NICER, Swift \citep{Mandel2025}, and IXPE \citep{Marra2025_IXPE} show that the X-ray flux peaked just before 2025 Apr 01 and around the time of these radio observations, indicating that the radio polarization evolves within a broader phase of multiwavelength variability. Given the potentially complicated and unresolved structure of this source, we cannot necessarily correlate changes in the intrinsic EVPA with substructure in the source, e.g., the jet's position angle, consistent with high-resolution observations of V404 Cyg with the VLBA \citep{MillerJones2019}.
        
    \subsection{A Knot Explanation for the Secondary Polarized Component}
        The secondary component was slightly more polarized than the primary component on 2025 Apr 06 as probed by the median posterior values in Table~\ref{tab:jointmodel}. However, the 68\% credible intervals for the two components significantly overlap, suggesting that the difference in the intrinsic polarization levels is not significant. The localized Faraday screen with an $\rm{RM}\approx-6200$ rad m$^{-2}$ reveals the line-of-sight magnetic field is oriented away from Earth (i.e., in the same direction as the GC screen).

        The large RM value in this secondary Faraday screen suggests a strong magnetic field, potentially within a jet, and/or a magnetic field largely along the line-of-sight. We take two approaches to test if this is a viable explanation. First, assuming this component cools only via synchrotron emission allows for an estimate of the (total) magnetic field strength. Using standard synchrotron relationships \citep[in][]{RybickiLightman1986,ERA2016}, and assuming an isotropic pitch angle, the estimated local magnetic field strength $B_{\rm{local}}$ and synchrotron cooling timescale $t_{\rm{cool}}$ are related by:
            \begin{equation}
                B_{\rm{local}} = 2 \left(\dfrac{t_{\rm{cool}}}{1\rm{~yr}}\right)^{-2/3}\left(\dfrac{\nu}{1\rm{~GHz}}\right)^{-1/3}~\rm{G}.
            \end{equation}
        If  $t_{\rm{cool}}=1$ day, estimated from the secondary component appearing on 2025 Apr 06 and disappearing on 2025 Apr 07, and $\nu=43$~GHz,  $B_{\rm{local}}\approx30$ gauss. A cooling timescale of 1 day is a lower bound on the true value, however, as this secondary component could have formed anytime after the end of the first observation (2025 Apr 04). If $t_{\rm{cool}}=3$ days, $B_{\rm{local}} \approx15$ gauss. Therefore, the synchrotron estimate is  consistent with a local magnetic field strength of 15--30 gauss. The cooling-timescale-based estimate does not account for higher-energy electrons that cool to the radio and emit (lengthening the required cooling time for the knot) and ignores any adiabatic cooling/expansion of the knot moving through the jet opening angle (biasing the magnetic field strength estimate to higher values).

        The second estimate comes from the analytic model developed for MAXI J1820+070  \citep{Zdziarski2022} and used by \citet{Hughes2023} for the 2015 radio outburst of V404 Cygni.  This model estimates the magnetic-field strength and electron number density ($n_e$) in a steady-state, conical jet. Following \citet{Hughes2023}, we adopt the fiducial jet parameters from \citet{Zdziarski2022} for MAXI J1820+070 in this computation. At 33~GHz, we estimate a total magnetic field strength $\approx$20 gauss and $n_e\approx6100$ cm$^{-3}$. Assuming the localized Faraday screen can be described by a one-zone model, the RM is related to the line-of-sight magnetic field strength ($B_{\parallel}$), electron number density, and thickness of the screen ($dl$) by $\textrm{RM}=n_e~B_{\parallel}~dl$. With $\left|\rm{RM}\right| = 6223$ rad m$^{-2}$, assuming $B_{\parallel} = 0.5$,  $B \approx10$ gauss, and $n_e=6100$ cm$^{-3}$ gives $dl \approx0.03$ AU. \citet{Hughes2023} derived a screen thickness of $0.06$ AU for V404 Cygni, assuming the observed RM was produced by a radio jet. 
        
        The two magnetic field-strength estimates and the Faraday screen depth are broadly consistent with the physical conditions in a polarized radio jet. These results suggest the secondary component formed as a short-lived knot within a jet and cooled via synchrotron radiation within  the jet's strong magnetic field. Depending on the optical depth of the potential knot, the intrinsic EVPA may trace the jet's orientation.
        
    \subsection{Radial Profile of the Rotation Measure in the Central Parsec}

        Before this work, only two objects in the central parsec of the Galaxy had measured RM values: \Sg\ and the GC magnetar (PSR J1745$-$2900) with RMs $\approx-5\times10^5$ \citep[e.g.,][]{Bower2003,Marrone2007,Bower2018} and $\approx-(\hbox{6--7})\times10^4$ rad m$^{-2}$ \citep{Eatough2013,Desvignes2018}, respectively. Farther out, however, the Galaxy within the inner few degrees has lower and/or positive RMs as probed by pulsars \citep[within a projected distance of $\sim$30 pc of \Sg;][]{Abbate2023} and background quasars \citep{Roy2008}, suggesting a variable and complicated magnetic configuration. Attempts to explain this large-to-small-scale change in the RM include local conditions \citep[i.e., higher magnetic field and particle densities in the GC;][]{Eatough2013}, Galactic structures \citep[i.e.,  spiral arms or the Galactic Bar,][]{Roy2008}, or  intervening line-of-sight sources \citep[such as \ion{H}{2} regions;][]{Sicheneder2017}. Moreover, the measured RMs of \Sg\ and PSR J1745$-$2900 appeared to temporally vary \citep[e.g.,][]{Bower2018,Desvignes2018}. \citet{Marrone2006} showed that the RM can be used to probe the accretion rate of \Sg. It is therefore crucial to understand the line-of-sight RM structure because treating an external RM as intrinsic to \Sg\ would affect the accretion rate measurement using this method. 

        \citet{Eatough2013} used measurements of the magnetic field ($B(r)$) and particle densities ($n(r)$) at arcminute-to-degree scales in the GC to argue that the observed RM of PSR J1745$-$2900 originates from a Faraday screen on $\sim$10 pc scales. They assumed that the magnetic field strength and particle density both vary as $r^{-1}$, yielding ${\rm{RM}}_{\rm{screen}}=B(r)\,n(r)\,r\propto r^{-1}$. This implies the GC screen substantially contributes to \Sg's measured RM, as ${\rm{RM}}_{\rm{screen}}$ increases towards \Sg. However, without a suitable secondary source, the local RM profile of the GC screen could not be tested.

        MAXI J1744 gives us this opportunity, as it is now the source with the largest projected radial separation from \Sg\ within the central parsec (projected radius of $18\farcs8$ corresponding to $\sim$0.75 pc at a distance of $8.178$ kpc; \citealt{GRAVITY2019}). Using the GC magnetar RM ($-66\,960 \pm 50$ rad m$^{-2}$ at projected radius of $\approx3\arcsec$, \citealt{Eatough2013}) as an anchor point with MAXI J1744's measured value ($-63\,606 \pm 850$ rad m$^{-2}$) yields $\rm{RM}\propto r^{-0.03 \pm 0.01}$ within the central parsec, which is extremely shallow and consistent with a flat RM screen on a scale of $\sim$1~pc. If no abrupt changes in the RM profile exist and the RM screen is azimuthally symmetric, our results support a similarly strong foreground RM $\approx-(\hbox{6--7})\times10^4$ rad m$^{-2}$ ($\approx$10\%) contribution to \Sg's measured RM\null. Consequently, the majority of \Sg's $\approx-5\times10^5$ rad m$^{-2}$ originates from the accretion flow rather than from an external source. 
        
        However, we cannot ascertain the true origin of the GC Faraday screen. For example, the best-fit range of \ion{H}{2}-region parameters capable of describing the large-scale scattering, dispersion, and Faraday characteristics toward the GC \citep{Sicheneder2017} has projected angular sizes of 1\arcmin--7\arcmin. Because MAXI J1744 lies within this range of projected radii from \Sg, our data are insensitive to spatial structures greater than its projected radius ($\approx$0\farcm3) and cannot rule out or constrain such origins. However, we can eliminate scenarios where the large RM present in the GC is \textit{only} locally produced given the nearly-flat RM profile, e.g., PSR J1745$-$2900 interacting with nearby ionized gas \citep{YusefZadeh2015}.


\section{Conclusions and Summary}\label{sec:conclusion}
We have presented the first linear polarization detection of the Galactic center transient MAXI J1744$-$294 with 33 and 43~GHz VLA observations in 2025 April.  Stokes $q$ and $u$ modeling shows that the linear polarization of MAXI J1744  varies on a $\sim$daily timescale.
\begin{enumerate}
    \item There is a persistent Faraday screen with rotation measure $=-63\,606^{+844}_{-861}$ rad m$^{-2}$, consistent with the Galactic center Faraday screen probed by the Galactic center magnetar PSR J1745$-$2900. This measured RM is the first direct evidence that MAXI J1744 resides within the Galactic center instead of in the foreground.
    \item There was a secondary polarized component on 2025 Apr 06 with a localized RM  $\approx-6200$ rad m$^{-2}$. This component was absent in polarization data taken before and after this date. Its nature is consistent with a short-lived knot in a putative radio jet with an $\approx$15--30 gauss magnetic field.
    \item The RMs for MAXI J1744 and PSR J1745$-$2900 are nearly equal. This supports the notion that most of \Sg's RM ($\sim-10^5$ rad m$^{-2}$) is intrinsic to the accretion flow rather than an unrelated source along the line-of-sight. 
\end{enumerate}

\appendix
\section{Polarization Conventions}\label{appx:pol_conventions}
    The Stokes parameters $I$, $Q$, and $U$ have uncertainties $\sigma_{\scriptscriptstyle I}$, $\sigma_{\scriptscriptstyle Q}$, and $\sigma_{\scriptscriptstyle U}$, respectively. The normalized Stokes parameters $q$ and $u$ therefore have uncertainties $\sigma_q$ and $\sigma_u$:
        \begin{align}
            \sigma_q &= \sqrt{\left(\dfrac{1}{I}\sigma_{\scriptscriptstyle Q}\right)^2 + \left(\dfrac{Q}{I^2}\sigma_{\scriptscriptstyle I}\right)^2},~\textrm{and}\\
            \sigma_u &= \sqrt{\left(\dfrac{1}{I}\sigma_{\scriptscriptstyle U}\right)^2 + \left(\dfrac{U}{I^2}\sigma_{\scriptscriptstyle I}\right)^2}.
        \end{align}
    The debiased linear polarization fraction $p_l$ is defined as:
        \begin{equation}
            p_l = \sqrt{q^2+u^2 - \sigma_{p_l}^2},
            \label{eq:pl}
        \end{equation}
    where $\sigma_{p_l}$ is the uncertainty on $p_l$:
        \begin{equation}
            \sigma_{p_l} = \dfrac{\sqrt{q^2\sigma_q^2 + u^2\sigma_u^2}}{\sqrt{q^2+u^2}}.
            \label{eq:error_pl}
        \end{equation}
    The debiased polarized intensity ($PI$) and uncertainty on $PI$ are equivalent to Equations \ref{eq:pl} and \ref{eq:error_pl}, respectively, under the change $\{q,u\}\rightarrow\{Q, U\}$.
    For completeness, the electric vector position angle (EVPA) in radians is defined by:
        \begin{equation}
            \psi = \dfrac{1}{2}\mathrm{arctan2}\left(\dfrac{u}{q}\right),
            \label{eq:evpa}
        \end{equation}
    with the uncertainty on the EVPA $\sigma_{\psi}$:
        \begin{equation}
            \sigma_{\psi} = \dfrac{1}{2}\dfrac{1}{q^2+u^2}\sqrt{q^2\sigma_u^2+u^2\sigma_q^2},
        \end{equation}
    where $\rm{arctan2}$ places the EVPA in the correct quadrant. Equivalent EVPA relationships using the normal Stokes parameters can again be determined with the change $\{q,u\}\rightarrow\{Q, U\}$.

\section{Estimating Residual Instrumental Polarization}\label{appx:dterm}

\begin{figure}[h]
    \centering
    \includegraphics[width=\linewidth]{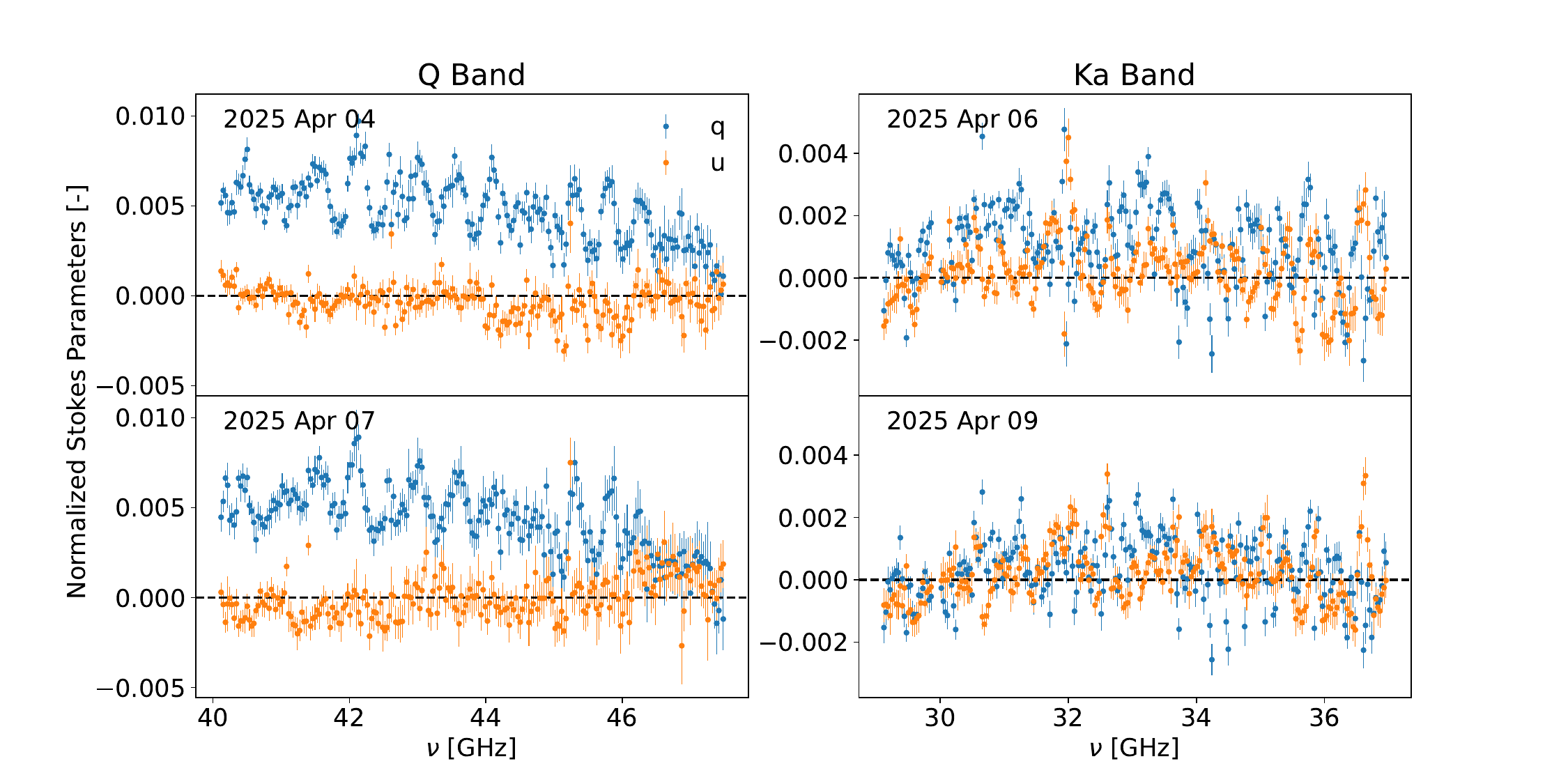}
    \caption{Normalized Stokes parameter spectropolarimetry of \Sg\ for all four observations. Blue and orange show $q$ and $u$ respectively as a function of observing frequency.  Each panel shows one night's results as labeled.}
    \label{fig:sgra_dterm}
\end{figure}

\Sg\ is well-known to be linearly unpolarized at radio frequencies \citep[e.g.,][]{Bower1999,Bower1999_lo}, and, as such, is a useful source to estimate the in-field residual instrumental polarization of the VLA observations given its large Stokes $I$ flux (often $\gtrsim$1.0 Jy). We applied the same aperture extraction methods as in Section \ref{sec:data} to obtain \Sg's Stokes $Q$ and $U$ spectropolarimetry except for centroiding the elliptical aperture in the Stokes $I$ plane instead of $PI$. The normalized Stokes parameters for \Sg\ are presented in Figure~\ref{appx:dterm}. Stokes $q$ and $u$ are centered near $0$ in the Ka-band observations and do not significantly vary between the two nights; this further confirms that the secondary polarized component observed on 2025 Apr 06 is not caused by instrumental calibration errors. The normalized Q-band Stokes $u$ parameter is also zero-centered in both observations, while the Stokes $q$ parameter is consistently offset by ${\approx}+0.005$ on 2025 Apr 04 and 07. This directly points toward an instrument effect as $u\approx0\implies\psi\approx0$ (Equation \ref{eq:evpa}), and, therefore, no RM ($\rm{RM}\equiv d\psi/d\lambda^2$) for \Sg, which lies behind the GC Faraday screen and is known to have a large (sub)millimeter-measured (absolute) RM value in excess of $10^5~\rm{rad~m}^{-2}$ \citep[e.g.,][]{Liu2016,Bower2018}. This ``corruption'' in Stokes $q$ may be traced back to an imperfect intrinsic polarization calibrator solution in \texttt{polcal} and/or an offset in the instrumental polarization leakage terms (see Table~1 of \citealt{Sault1996} for the circularly-polarized basis) and consistent with the nominal $\lesssim$0.5\% instrumental polarization accuracy for point sources\footnote{\url{https://science.nrao.edu/facilities/vla/docs/manuals/oss/performance/polarization}}. These spectropolarimetry results suggest residual instrumental polarizations of $\sim$0.2\% and $\sim$0.5\% in Ka and Q bands, respectively. Because we followed the VLA setup for polarimetric observations by fixing the absolute polarization angle to a standard source (in this case, J1331+3030), we expect no systematic issues in the EVPA between the Q and Ka bands.  MAXI J1744's measured intrinsic linear polarization in Q band ranges from $\hbox{4--10}\%\gg0.5\%$, implying no bias from residual instrumental polarization is dominant in the observation.

\section{2025 Apr 06 Flux Correction}\label{appx:fluxcorr}
    \begin{figure}[h]
        \centering
        \includegraphics[width=\linewidth]{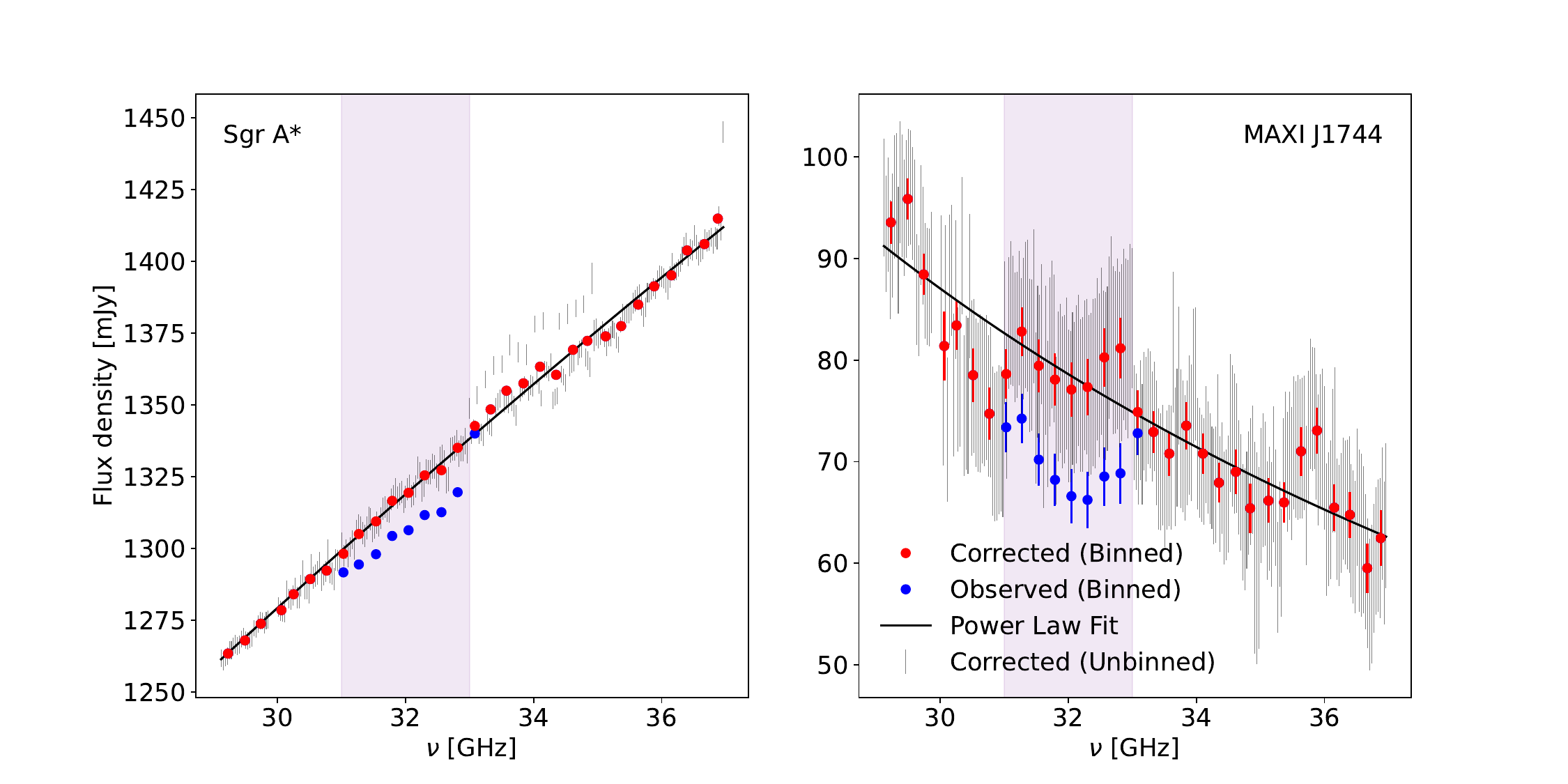}
        \caption{Binned observed (blue) and corrected (red) SEDs of \Sg\ and MAXI J1744, left and right panels, respectively, on 2025 Apr 06. The native resolution corrected SEDs are shown as gray points, and the best-fit power law to the corrected data is shown as a black line. The light purple shaded area denotes the range of frequencies where the correction was applied.}
        \label{fig:flux_fix}
    \end{figure}

The 31--33~GHz baseband flux density on 2025 Apr 06 was dimmer by $\approx$10 mJy relative to the rest of the Ka band observations for MAXI J1744. This issue was also present on the same night for \Sg, J1733$-$1304, and J1744$-$3116, potentially tracing back to an issue during flux bootstrapping in the VLA pipeline. Attempts to locate the origin of this shift in the calibrated data were unfruitful, as no antenna, baseline, time range, nor correlation appeared inconsistent with the rest of the data. We corrected the flux in the 31--33~GHz frequency range by assuming the SED of \Sg\ can be described by a power law across the full Ka band range. We fit the 29.1--31.0 and 33.0--37.0~GHz range of \Sg's SED with a power law and determined the residual (model $-$ observed) between 31 and 33~GHz. We then obtain a best-fit linear model to the residuals, which minimizes the measurement noise in the residuals, to calculate the additive offset at each frequency bin \textit{at the phase center} (i.e., the flux bootstrapping occurs with non-primary-beam corrected data). To apply corrections to the spectrum of MAXI J1744, we multiplied the flux corrections by the primary beam response ($\approx0.85$ at the location of the source) and applied these values to the observed spectra for both sources. Figure~\ref{fig:flux_fix} shows the SED of \Sg\ and MAXI J1744 before and after  flux correction, which largely recovers the power-law-like SED for both sources. After correction, MAXI J1744's spectrum still has lower flux densities near $\approx$30~GHz, but we did not correct these data further. The best-fit power law does not appear to be severely affected by this residual flux bootstrapping issue regardless.

\section{Model Fit Parameters}\label{appx:results}
    \input{qufit_table}

    The most basic polarization model in \texttt{RM-Tools} (\texttt{m1}) takes the form:
        \begin{equation}
            p_{\texttt{m1}}(\lambda)=p_0e^{2i\left(\psi_0+\phi\lambda^2\right)},
            \label{eq:m1}
        \end{equation}
    where $p_0$ and $\psi_0$ are the intrinsic polarization fraction and EVPA, respectively, at a wavelength $\lambda=0$. $\phi$ is the Faraday depth of the intervening Faraday screen. This model describes a single polarized component passing through a non-emitting Faraday screen.

    We also fit each epoch with a secondary polarization model (\texttt{RM-Tools} model ``\texttt{m11}''), consisting of two disconnected polarized components within a single beam, each with its own Faraday depth, intrinsic polarization fraction, and intrinsic polarization angle:
        \begin{equation}
            p_{\texttt{m11}}(\lambda)=p_{0,1}e^{2i\left(\psi_{0,1}+\phi_1\lambda^2\right)} + p_{0,2}e^{2i\left(\psi_{0,2}+\phi_2\lambda^2\right)},
            \label{eq:m11}
        \end{equation}
    where the individual values have the same meaning as above. In this form, $\phi_1 = \phi_{\rm{GC}}$ and $\phi_2 = \phi_{\rm{GC}}+\phi_{\rm{local}}$ (see Equation~\ref{eq:apr6}). In Equations~\ref{eq:m1} and \ref{eq:m11} above, the normalized Stokes $q$ and $u$ products are the real and imaginary parts of $p\left(\lambda\right)$, respectively.

    The best-fit parameters for these two models are listed in Table~\ref{tab:qumodels} along with three goodness-of-fit measures calculated by \texttt{RM-Tools}: $\chi^2$, the Bayesian Information Criterion (BIC), and the natural logarithm of the Bayesian evidence.
    \texttt{RM-Tools} uses the nested sampler package \texttt{dynesty} \citep{dynesty} to calculate these measures.
For the BIC, only differences are significant, and we adopted the standard $\left|\Delta\rm{BIC}\right|\equiv\left|\rm{BIC}_2-\rm{BIC}_1\right|>10$ value, which signifies model 2 ($\rm{BIC}_2$) better represents the data. We chose the better-fitting model as the one preferred by at least two goodness-of-fit parameters. For 2025 Apr 06, we did the fitting twice for each model: once with all data and once without the 31--33~GHz data. This explicitly ensures that the preferred two-component polarized model on this night is not influenced by the flux scaling technique explained in Appendix~\ref{appx:fluxcorr}. In both cases, the two-component model is preferred over a single component.

\begin{acknowledgments}
We thank St\'{e}phane Corbel, Noa Grollimund, and Alexander Philippov for discussions about MAXI J1744 early in this paper's analysis, and Shifa Mandel for comments on a draft of this work. We are indebted to the VLA scheduling and operations teams who made these observations possible and to the anonymous referee whose comments helped strengthen this work. The National Radio Astronomy Observatory is a facility of the National Science Foundation operated under cooperative agreement by Associated Universities, Inc. J.M.M is supported by an NSF Astronomy and Astrophysics Postdoctoral Fellowship under award AST-2401752. S.D.v.F. gratefully acknowledges the support of the Alexander von Humboldt Foundation through a Feodor Lynen Fellowship, and thanks CITA for their hospitality and collaboration. Some of the analysis in this paper were conducted on the Smithsonian High Performance Cluster (SI/HPC), Smithsonian Institution (\url{https://doi.org/10.25572/SIHPC}). B.R. is supported by the Natural Sciences \& Engineering Research Council of Canada (NSERC) [funding reference number 568580] and acknowledges support by a grant from the Simons Foundation (MP-SCMPS-00001470). B.R., D.H., N.M.F., Z.S., and M.B. acknowledge support from the Canadian Space Agency under funding number 23JWGO2A01. This research was supported in part by grant NSF PHY-2309135 to the Kavli Institute for Theoretical Physics (KITP). N.M.F. and D.H. acknowledge funding from the Natural Sciences and Engineering Research Council of Canada (NSERC) Discovery Grant program and the Canada Research Chairs (CRC) program. N.M.F. acknowledges funding from the Fondes de Recherche Nature et Technologies (FRQNT) Doctoral research scholarship. N.M.F. and D.H. acknowledge support from the Canadian New Frontiers in Research Fund (NFRF) -- Explorations program and the Trottier Space Institute at McGill. The authors acknowledge support from the Centre de recherche en astrophysique du Québec, un regroupement stratégique du FRQNT. S.M. is supported by a European Research Council Synergy Grant “Blackholistic” (grant 10107164). M.B. acknowledges support from the Natural Sciences and Engineering Research Council of Canada's Banting Postdoctoral Fellowship Program (CIHR AWARD BPF 200617-267964).

\end{acknowledgments}




%
\facilities{VLA}

\software{ CASA \citep{CASA2022}, astropy \citep{astropy:2013, astropy:2018, astropy:2022},
          \texttt{dynesty} \citep{dynesty}, \texttt{RM-Tools} \citep{RMTools2020}}



\bibliography{sample701}{}
\bibliographystyle{aasjournalv7}



\end{document}

%% file: qufit_table.tex
\begin{deluxetable*}{l|c|lcc|ccc}
            \tabletypesize{\footnotesize}
            \tablewidth{\linewidth} 
            \tablenum{4}
            \tablecaption{Stokes $q$ and $u$ model fits for MAXI J1744.\label{tab:qumodels}}
            \tablehead{\colhead{Date} & \colhead{Model} & \colhead{Parameter} & \colhead{Prior} & \colhead{Posterior (68\% CI)} & \colhead{$\chi^2$/DOF} & \colhead{BIC} & \colhead{ln Evidence}}
            \startdata
                2025 Apr 04  & \texttt{m1}$\bigstar$ & $p_0~[\%]$ & $\mathcal{U}[0, 100]$ & $9.83^{+0.16}_{-0.15}$ & & & \\
                             &             & $\phi$ [rad m$^{-2}$] & $\mathcal{U}[-110000, 110000]$ & $-63402^{+1856}_{-1705}$ & $\mathbf{303.80/488}$ & $\mathbf{-1954.79}$ & $972.21 \pm 0.30$\\
                             &             & $\psi_0~[^\circ]$ & $\mathcal{U}[0, 180]$ & $39.29^{+4.65}_{-5.15}$ & & & \\
                \cline{2-8}
                             & \texttt{m11}\tablenotemark{a} & $p_{0,1}~[\%]$          & $\mathcal{U}[0, 100]$ & $9.84^{+2.14}_{-1.46}$ & & & \\
                             &              & $p_{0,2}~[\%]$          & $\mathcal{U}[0, 100]$ & $3.18^{+3.05}_{-1.42}$& & & \\
                             &              & $\phi_1$ [rad m$^{-2}$] & $\mathcal{U}[-110000, 110000]$ & $-63630^{+5719}_{-5400}$ & $750.88/485$ & $-1960.61$ & $\mathbf{980.72\pm0.33}$\\
                             &              & $\phi_2$ [rad m$^{-2}$] & $\mathcal{U}[-110000, 110000]$ & $-41748^{+29216}_{-43667}$ &  & & \\
                             &              & $\psi_{0,1}~[^\circ]$   & $\mathcal{U}[0, 180]$ & $43.03^{+16.02}_{-16.19}$ & & & \\
                             &              & $\psi_{0,2}~[^\circ]$   & $\mathcal{U}[0, 180]$ & $67.37^{+81.57}_{-42.40}$ & & & \\\hline\hline
                2025 Apr 06  & \multicolumn{7}{c}{All data included in $q$--$u$ fitting.}\\
                 \cline{2-8}
               & \texttt{m1} & $p_0~[\%]$ & $\mathcal{U}[0, 100]$ & $4.85^{+0.05}_{-0.05}$ & & & \\
                             &             & $\phi$ [rad m$^{-2}$] & $\mathcal{U}[-110000, 110000]$ & $-74582^{+419}_{-437}$ & $738.30/480$ & $-2609.25$ & $1297.00 \pm 0.32$\\
                             &             & $\psi_0~[^\circ]$ & $\mathcal{U}[0, 180]$ & $75.36^{+2.25}_{-2.10}$ & & & \\
                \cline{2-8}
                             & \texttt{m11}$\bigstar$ & $p_{0,1}~[\%]$          & $\mathcal{U}[0, 100]$ &  $9.99^{+3.93}_{-2.07}$  & & & \\
                             &              & $p_{0,2}~[\%]$          & $\mathcal{U}[0, 100]$ &  $6.63^{+3.85}_{-2.05}$ & & & \\
                             &              & $\phi_1$ [rad m$^{-2}$] & $\mathcal{U}[-110000, 110000]$ & $-69767^{+1867}_{-2105}$ & $\mathbf{338.00/477}$ & $\mathbf{-2965.25}$ & $\mathbf{1475.17\pm0.40}$ \\
                             &              & $\phi_2$ [rad m$^{-2}$] & $\mathcal{U}[-110000, 110000]$ & $-62047^{+3095}_{-2744}$  & & & \\
                             &              & $\psi_{0,1}~[^\circ]$   & $\mathcal{U}[0, 180]$ &  $34.27^{+9.84}_{-8.99}$ & & & \\
                             &              & $\psi_{0,2}~[^\circ]$   & $\mathcal{U}[0, 180]$ &  $72.58^{+11.86}_{-15.36}$&  & & \\
                \cline{2-8}
                & \multicolumn{7}{c}{31--33 GHz data excluded from $q$--$u$ fitting.}\\
                \cline{2-8}
                             & \texttt{m1} & $p_0~[\%]$ & $\mathcal{U}[0, 100]$ & $4.88^{+0.06}_{-0.06}$ & & & \\
                             &             & $\phi$ [rad m$^{-2}$] & $\mathcal{U}[-110000, 110000]$ & $-74436^{+417}_{-431}$ & $612.89/354$ & $-1836.14$ & $909.95 \pm 0.32$\\
                             &             & $\psi_0~[^\circ]$ & $\mathcal{U}[0, 180]$ & $74.05^{+2.26}_{-2.17}$ & & & \\
                \cline{2-8}
                             & \texttt{m11}$\bigstar$ & $p_{0,1}~[\%]$          & $\mathcal{U}[0, 100]$ &  $7.16^{+4.06}_{-1.68}$  & & & \\
                             &              & $p_{0,2}~[\%]$          & $\mathcal{U}[0, 100]$ &  $5.43^{+3.97}_{-1.55}$ & & & \\
                             &              & $\phi_1$ [rad m$^{-2}$] & $\mathcal{U}[-110000, 110000]$ & $-73689^{+2770}_{-2543}$ & $\mathbf{214.02/351}$ & $\mathbf{-2192.23}$ & $\mathbf{1088.33\pm0.39}$ \\
                             &              & $\phi_2$ [rad m$^{-2}$] & $\mathcal{U}[-110000, 110000]$ & $-81978^{+2567}_{-3717}$  & & & \\
                             &              & $\psi_{0,1}~[^\circ]$   & $\mathcal{U}[0, 180]$ &  $94.24^{+13.72}_{-14.02}$ & & & \\
                             &              & $\psi_{0,2}~[^\circ]$   & $\mathcal{U}[0, 180]$ &  $70.31^{+17.13}_{-15.49}$&  & & \\\hline\hline
                2025 Apr 07  & \texttt{m1}$\bigstar$ & $p_0~[\%]$ & $\mathcal{U}[0, 100]$ & $3.63^{+0.11}_{-0.11}$ & & & \\
                             &             & $\phi$ [rad m$^{-2}$] & $\mathcal{U}[-110000, 110000]$ & $-63790^{+3303}_{-3409}$ & $\mathbf{261.80/488}$ & $\mathbf{-2355.35}$ & $\mathbf{1173.36 \pm 0.29}$\\
                             &             & $\psi_0~[^\circ]$ & $\mathcal{U}[0, 180]$ & $62.64^{+9.16}_{-9.51}$ & & & \\
                \cline{2-8}
                             & \texttt{m11} & $p_{0,1}~[\%]$          & $\mathcal{U}[0, 100]$ & $4.07^{+3.09}_{-0.06}$ & & & \\
                             &              & $p_{0,2}~[\%]$          & $\mathcal{U}[0, 100]$ & $0.95^{+2.83}_{-0.72}$& & & \\
                             &              & $\phi_1$ [rad m$^{-2}$] & $\mathcal{U}[-110000, 110000]$ & $-65797^{+5909}_{-9832}$ & $400.41/485$ & $-2343.36$ & $1171.13\pm0.32$ \\
                             &              & $\phi_2$ [rad m$^{-2}$] & $\mathcal{U}[-110000, 110000]$ & $-70581^{+40169}_{-27866}$ & & & \\
                             &              & $\psi_{0,1}~[^\circ]$   & $\mathcal{U}[0, 180]$ & $68.68^{+25.47}_{-17.90}$ & & & \\
                             &              & $\psi_{0,2}~[^\circ]$   & $\mathcal{U}[0, 180]$ & $62.92^{+65.59}_{-39.66}$& & & \\\hline \hline
                2025 Apr 09  & \texttt{m1}$\bigstar$ & $p_0~[\%]$ & $\mathcal{U}[0, 100]$ & $1.54^{+0.04}_{-0.04}$ & & & \\
                             &             & $\phi$ [rad m$^{-2}$] & $\mathcal{U}[-110000, 110000]$ & $-63873^{+1039}_{-1205}$ & $\mathbf{261.78/474}$ & $\mathbf{-3306.87}$ & $1646.38 \pm 0.32$\\
                             &             & $\psi_0~[^\circ]$ & $\mathcal{U}[0, 180]$ & $63.47^{+6.13}_{-5.23}$ & & & \\
                \cline{2-8}
                             & \texttt{m11} & $p_{0,1}~[\%]$          & $\mathcal{U}[0, 100]$ & $1.49^{+0.06}_{-0.16}$ & & & \\
                             &              & $p_{0,2}~[\%]$          & $\mathcal{U}[0, 100]$ & $0.24^{+0.15}_{-0.07}$& & & \\
                             &              & $\phi_1$ [rad m$^{-2}$] & $\mathcal{U}[-110000, 110000]$ & $-68468^{+2723}_{-4235}$ & $317.07/471$ & $-3308.82$ & $\mathbf{1648.87\pm0.38}$\\
                             &              & $\phi_2$ [rad m$^{-2}$] & $\mathcal{U}[-110000, 110000]$ & $5034^{+18400}_{-20272}$ & & & \\
                             &              & $\psi_{0,1}~[^\circ]$   & $\mathcal{U}[0, 180]$ & $85.91^{+21.11}_{-13.03}$ & & & \\
                             &              & $\psi_{0,2}~[^\circ]$   & $\mathcal{U}[0, 180]$ & $97.41^{+56.43}_{-66.63}$& & & \\\hline\hline
            \enddata
            \tablecomments{Bold numbers in the three rightmost columns indicate the model (\texttt{m1} or \texttt{m11}) with the better goodness-of-fit value.}
            \tablenotetext{a}{Two constraints are used for the \texttt{m11} fits: 1) $p_{0,1}+p_{0,2}\leq100\%$, 2) $\Delta p_0=p_{0,1}-p_{0,2}\in[0, 100]\%$ (i.e., so $p_{0, 1}\geq p_{0, 2}$).}
            \tablenotetext{\bigstar}{~Chosen best-fit model for each observation.}
        \end{deluxetable*}

%% file: sample701.bib
@ARTICLE{CASA2022,
       author = {{CASA Team} and {Bean}, Ben and {Bhatnagar}, Sanjay and {Castro}, Sandra and {Donovan Meyer}, Jennifer and {Emonts}, Bjorn and {Garcia}, Enrique and {Garwood}, Robert and {Golap}, Kumar and {Gonzalez Villalba}, Justo and {Harris}, Pamela and {Hayashi}, Yohei and {Hoskins}, Josh and {Hsieh}, Mingyu and {Jagannathan}, Preshanth and {Kawasaki}, Wataru and {Keimpema}, Aard and {Kettenis}, Mark and {Lopez}, Jorge and {Marvil}, Joshua and {Masters}, Joseph and {McNichols}, Andrew and {Mehringer}, David and {Miel}, Renaud and {Moellenbrock}, George and {Montesino}, Federico and {Nakazato}, Takeshi and {Ott}, Juergen and {Petry}, Dirk and {Pokorny}, Martin and {Raba}, Ryan and {Rau}, Urvashi and {Schiebel}, Darrell and {Schweighart}, Neal and {Sekhar}, Srikrishna and {Shimada}, Kazuhiko and {Small}, Des and {Steeb}, Jan-Willem and {Sugimoto}, Kanako and {Suoranta}, Ville and {Tsutsumi}, Takahiro and {van Bemmel}, Ilse M. and {Verkouter}, Marjolein and {Wells}, Akeem and {Xiong}, Wei and {Szomoru}, Arpad and {Griffith}, Morgan and {Glendenning}, Brian and {Kern}, Jeff},
        title = "{CASA, the Common Astronomy Software Applications for Radio Astronomy}",
      journal = {\pasp},
         year = 2022,
        month = nov,
       volume = {134},
       number = {1041},
          eid = {114501},
        pages = {114501},
          doi = {10.1088/1538-3873/ac9642},
archivePrefix = {arXiv},
       eprint = {2210.02276},
 primaryClass = {astro-ph.IM},
       adsurl = {https://ui.adsabs.harvard.edu/abs/2022PASP..134k4501C}
}

@ARTICLE{automultithresh,
       author = {{Kepley}, Amanda A. and {Tsutsumi}, Takahiro and {Brogan}, Crystal L. and {Indebetouw}, Remy and {Yoon}, Ilsang and {Mason}, Brian and {Donovan Meyer}, Jennifer},
        title = "{Auto-multithresh: A General Purpose Automasking Algorithm}",
      journal = {\pasp},
         year = 2020,
        month = feb,
       volume = {132},
       number = {1008},
          eid = {024505},
        pages = {024505},
          doi = {10.1088/1538-3873/ab5e14},
archivePrefix = {arXiv},
       eprint = {1912.04970},
 primaryClass = {astro-ph.IM},
       adsurl = {https://ui.adsabs.harvard.edu/abs/2020PASP..132b4505K}
}

@ARTICLE{Bower1999,
       author = {{Bower}, Geoffrey C. and {Backer}, Donald C. and {Zhao}, Jun-Hui and {Goss}, Miller and {Falcke}, Heino},
        title = "{The Linear Polarization of Sagittarius A*. I. VLA Spectropolarimetry at 4.8 and 8.4 GHZ}",
      journal = {\apj},
         year = 1999,
        month = aug,
       volume = {521},
       number = {2},
        pages = {582-586},
          doi = {10.1086/307592},
archivePrefix = {arXiv},
       eprint = {astro-ph/9904091},
 primaryClass = {astro-ph},
       adsurl = {https://ui.adsabs.harvard.edu/abs/1999ApJ...521..582B}
}

@ARTICLE{2025ATel_ninjasat_watanabe,
       author = {{Watanabe}, S. and {Aoyama}, A. and {Takeda}, T. and {Iwata}, S. and {Ota}, N. and {Yamasaki}, K. and {Takahashi}, T. and {Jujo}, A. and {Mihara}, T. and {Iwakiri}, W. and {Tamagawa}, T. and {Enoto}, T. and {Kitaguchi}, T. and {Kato}, Y. and {Hu}, C. -P.},
        title = "{NinjaSat Follow-up Observation of the X-ray Transient in the Galactic Center Region}",
      journal = {The Astronomer's Telegram},
         year = 2025,
        month = jan,
       volume = {17009},
        pages = {1},
       adsurl = {https://ui.adsabs.harvard.edu/abs/2025ATel17009....1W}
}

@ARTICLE{2025ATel_nicer_jaisawal,
       author = {{Jaisawal}, Gaurava K. and {Steiner}, James F. and {Strohmayer}, T.~E. and {Koenig}, Ole and {Altamirano}, Diego and {Arzoumanian}, Z. and {Gendreau}, K.~C. and {Sanna}, A. and {Draghis}, Paul and {Buisson}, Douglas},
        title = "{NICER Observation of the Galactic Center Transient MAXI J1744-294}",
      journal = {The Astronomer's Telegram},
         year = 2025,
        month = feb,
       volume = {17040},
        pages = {1},
       adsurl = {https://ui.adsabs.harvard.edu/abs/2025ATel17040....1J}
}

@ARTICLE{2025ATel_einstein_wang,
       author = {{Wang}, Yilong and {Coti Zelati}, Francesco and {Rea}, Nanda and {Jin}, Chichuan and {Marino}, Alessio and {Hu}, Jingwei and {Zhang}, Minghao and {Lian}, Tianying and {Dai}, Cuiyuan and {Zhao}, Yaqi and {Zhang}, Wenda and {Kuulkers}, Erik and {Yuan}, Weimin and {Guillot}, Sebastien and {Cheng}, Huaqing and {Zhang}, Guobao and {Kong}, Albert and {Li}, Zhaosheng},
        title = "{Einstein Probe observations of the new Galactic transient MAXI J1744-294}",
      journal = {The Astronomer's Telegram},
         year = 2025,
        month = mar,
       volume = {17068},
        pages = {1},
       adsurl = {https://ui.adsabs.harvard.edu/abs/2025ATel17068....1W}
}

@ARTICLE{2025ATel_xrism_mandel,
       author = {{Mandel}, S. and {Mori}, K. and {Hailey}, C. and {Bachetti}, M. and {Degenaar}, N. and {Draghis}, P. and {Grindlay}, J. and {Hong}, J. and {Ponti}, G. and {Reynolds}, M. and {Tomsick}, J. and {Wijnands}, R. and {Fukushima}, K. and {Hayashi}, K. and {Kanemaru}, Y. and {Ogawa}, S. and {Yoshida}, T. and {Iizuka}, R. and {Baluta}, C. and {Terada}, Y. and {Costantini}, E. and {Guainazzi}, M. and {Kelley}, R. and {Matsushita}, K. and {Petre}, R. and {Williams}, B. and {Yamaguchi}, H. and {Watanabe}, S. and {Tashiro}, M.},
        title = "{XRISM follow-up observation of a new transient MAXI J1744-294 at the Galactic central region}",
      journal = {The Astronomer's Telegram},
         year = 2025,
        month = mar,
       volume = {17063},
        pages = {1},
       adsurl = {https://ui.adsabs.harvard.edu/abs/2025ATel17063....1M}
}

@ARTICLE{2025ATel_chandra_mandel,
       author = {{Mandel}, S. and {Levin}, B. and {Mori}, K. and {Hailey}, C. and {Bachetti}, M. and {Degenaar}, N. and {Draghis}, P. and {Grindlay}, J. and {Hong}, J. and {Nynka}, M. and {Parra}, M. and {Ponti}, G. and {Reynolds}, M. and {Shidatsu}, M. and {Tomsick}, J. and {Wijnands}, R.},
        title = "{Chandra position of the new Galactic center transient MAXI J1744-294}",
      journal = {The Astronomer's Telegram},
         year = 2025,
        month = mar,
       volume = {17087},
        pages = {1},
       adsurl = {https://ui.adsabs.harvard.edu/abs/2025ATel17087....1M}
}

@ARTICLE{2025ATel_meerkat_grollimund,
       author = {{Grollimund}, Noa and {Corbel}, Stephane and {Bahramian}, Arash and {Carotenuto}, Francesco and {Heywood}, Ian and {Cowie}, Fraser and {Hughes}, Andrew and {Fender}, Rob and {Motta}, Sara},
        title = "{Discovery of the radio counterpart of MAXI J1744-294 with MeerKAT}",
      journal = {The Astronomer's Telegram},
         year = 2025,
        month = feb,
       volume = {17045},
        pages = {1},
       adsurl = {https://ui.adsabs.harvard.edu/abs/2025ATel17045....1G}
}

@ARTICLE{Eatough2013,
       author = {{Eatough}, R.~P. and {Falcke}, H. and {Karuppusamy}, R. and {Lee}, K.~J. and {Champion}, D.~J. and {Keane}, E.~F. and {Desvignes}, G. and {Schnitzeler}, D.~H.~F.~M. and {Spitler}, L.~G. and {Kramer}, M. and {Klein}, B. and {Bassa}, C. and {Bower}, G.~C. and {Brunthaler}, A. and {Cognard}, I. and {Deller}, A.~T. and {Demorest}, P.~B. and {Freire}, P.~C.~C. and {Kraus}, A. and {Lyne}, A.~G. and {Noutsos}, A. and {Stappers}, B. and {Wex}, N.},
        title = "{A strong magnetic field around the supermassive black hole at the centre of the Galaxy}",
      journal = {\nat},
         year = 2013,
        month = sep,
       volume = {501},
       number = {7467},
        pages = {391-394},
          doi = {10.1038/nature12499},
archivePrefix = {arXiv},
       eprint = {1308.3147},
 primaryClass = {astro-ph.GA},
       adsurl = {https://ui.adsabs.harvard.edu/abs/2013Natur.501..391E}
}

@ARTICLE{2025ATel_maxi_kudo,
       author = {{Kudo}, Y. and {Negoro}, H. and {Nakajima}, M. and {Shibui}, H. and {Takagi}, K. and {Takahashi}, H. and {Tatano}, K. and {Nishio}, H. and {Mihara}, T. and {Kawamuro}, T. and {Yamada}, S. and {Wang}, S. and {Tamagawa}, T. and {Kawai}, N. and {Matsuoka}, M. and {Sakamoto}, T. and {Serino}, M. and {Sugita}, S. and {Kawakubo}, Y. and {Hiramatsu}, H. and {Nishikawa}, H. and {Kondo}, Y. and {Sasao}, S. and {Yoshida}, A. and {Tsuboi}, Y. and {Sugai}, H. and {Nagashima}, N. and {Shidatsu}, M. and {Niida}, Y. and {Takahashi}, I. and {Niwano}, M. and {Higuchi}, N. and {Yatsu}, Y. and {Nakahira}, S. and {Ueno}, S. and {Tomida}, H. and {Ishikawa}, M. and {Ogawa}, S. and {Kurihara}, M. and {Ueda}, Y. and {Okada}, Y. and {Fujiwara}, K. and {Yamauchi}, M. and {Otsuki}, Y. and {Hasegawa}, T. and {Nishio}, M. and {Yamaoka}, K. and {Sugizaki}, M. and {Iwakiri}, W.},
        title = "{MAXI/GSC detection of a bright X-ray outburst from KS 1741-293 or a new hard X-ray transient MAXI J1744-294 near the galactic center}",
      journal = {The Astronomer's Telegram},
         year = 2025,
        month = jan,
       volume = {16975},
        pages = {1},
       adsurl = {https://ui.adsabs.harvard.edu/abs/2025ATel16975....1K}
}

@ARTICLE{2025ATel_maxi_nakajima,
       author = {{Nakajima}, M. and {Negoro}, H. and {Kudo}, Y. and {Mihara}, T. and {Shibui}, H. and {Takagi}, K. and {Takahashi}, H. and {Tatano}, K. and {Nishio}, H. and {Kawamuro}, T. and {Yamada}, S. and {Wang}, S. and {Tamagawa}, T. and {Kawai}, N. and {Matsuoka}, M. and {Sakamoto}, T. and {Serino}, M. and {Sugita}, S. and {Kawakubo}, Y. and {Hiramatsu}, H. and {Nishikawa}, H. and {Kondo}, Y. and {Sasao}, S. and {Yoshida}, A. and {Tsuboi}, Y. and {Sugai}, H. and {Nagashima}, N. and {Shidatsu}, M. and {Niida}, Y. and {Takahashi}, I. and {Niwano}, M. and {Higuchi}, N. and {Yatsu}, Y. and {Nakahira}, S. and {Ueno}, S. and {Tomida}, H. and {Ishikawa}, M. and {Ogawa}, S. and {Kurihara}, M. and {Ueda}, Y. and {Okada}, Y. and {Fujiwara}, K. and {Yamauchi}, M. and {Otsuki}, Y. and {Hasegawa}, T. and {Nishio}, M. and {Yamaoka}, K. and {Sugizaki}, M. and {Iwakiri}, W.},
        title = "{MAXI/GSC observation of the bright X-ray outburst from the Galactic center region}",
      journal = {The Astronomer's Telegram},
         year = 2025,
        month = jan,
       volume = {16983},
        pages = {1},
       adsurl = {https://ui.adsabs.harvard.edu/abs/2025ATel16983....1N}
}

@ARTICLE{2025ATel_swift_Heinke,
       author = {{Heinke}, C.~O. and {Nakajima}, M. and {Kudo}, Y. and {Negoro}, H. and {Bahramian}, A. and {Miller-Jones}, J. and {Degenaar}, N. and {Maccarone}, T. and {Yusef-Zadeh}, F. and {Sivakoff}, G.~R. and {Ridder}, M.},
        title = "{Swift/XRT position of MAXI J1744-294 near Sgr A*}",
      journal = {The Astronomer's Telegram},
         year = 2025,
        month = feb,
       volume = {17010},
        pages = {1},
       adsurl = {https://ui.adsabs.harvard.edu/abs/2025ATel17010....1H}
}

@ARTICLE{2025ATel17174_vla_michail,
       author = {{Michail}, J. M. and {von Fellenberg}, S. and {Haggard}, D. and {Fazio}, G. and {Ford}, N. and {Hora}, J. and {Nowak}, M. and {Neilsen}, J. and {Philippov}, A. and {Ripperda}, B. and {Roychowdhury}, T. and {Sanchez-Maes}, S. and {Smith}, H.~A. and {Sumners}, Z. and {Willner}, S. and {Seefeldt-Gail}, B. and {Witzel}, G.},
        title = "{Quick Radio Brightening of Galactic Center Transient MAXI J1744-294 at 33 and 43 GHz}",
      journal = {The Astronomer's Telegram},
         year = 2025,
        month = may,
       volume = {17174},
        pages = {1},
       adsurl = {https://ui.adsabs.harvard.edu/abs/2025ATel17174....1M}
}

@ARTICLE{Mandel2025,
       author = {{Mandel}, Shifra and {Mori}, Kaya and {Ciurlo}, Anna and {Draghis}, Paul A. and {Grollimund}, Noa and {Jaisawal}, Gaurava K. and {Jin}, Chichuan and {Levin}, Benjamin and {Marra}, Lorenzo and {Miao}, Eric and {Parra}, Maxime and {Reynolds}, Mark and {Granados}, Sean A. and {Bachetti}, Matteo and {Capitanio}, Fiamma and {Degenaar}, Nathalie and {Hailey}, Charles J. and {Hong}, JaeSub and {Motta}, Sara and {Ponti}, Gabriele and {Shara}, Michael M. and {Shidatsu}, Megumi and {Tomsick}, John A. and {Campbell}, Randall and {Corbel}, St{\'e}phane and {Fender}, Rob and {Ghez}, Andrea and {Grindlay}, Jonathan and {Hosek}, Jr., Matthew W. and {Matsunaga}, Kai and {Miku{\v{s}}incov{\'a}}, Romana and {Nynka}, Melania and {Sanger-Johnson}, Grace and {Stel}, Giovanni and {Tarana}, Antonella and {Wijnands}, Rudy and {Zhang}, Shuo},
        title = "{A multiwavelength study of the new Galactic center black hole candidate MAXI J1744-294}",
      journal = {arXiv e-prints},
         year = 2025,
        month = sep,
          eid = {arXiv:2509.14465},
        pages = {arXiv:2509.14465},
          doi = {10.48550/arXiv.2509.14465},
archivePrefix = {arXiv},
       eprint = {2509.14465},
 primaryClass = {astro-ph.HE},
       adsurl = {https://ui.adsabs.harvard.edu/abs/2025arXiv250914465M}
}

@ARTICLE{Marra2025_IXPE,
       author = {{Marra}, Lorenzo and {Miku{\v{s}}incov{\'a}}, Romana and {Vincentelli}, Federico M. and {Capitanio}, Fiamma and {Del Santo}, Melania and {Fabiani}, Sergio and {Mandel}, Shifra and {Muleri}, Fabio and {Parra}, Maxime and {Soffitta}, Paolo and {Tarana}, Antonella and {Baglio}, M. Cristina and {Bianchi}, Stefano and {Costa}, Enrico and {D'A{\`\i}}, Antonino and {De Marco}, Barbara and {Dov{\v{c}}iak}, Michal and {Elvezia Gianolli}, Vittoria and {Gnarini}, Andrea and {Gupta}, Maitrayee and {Ingram}, Adam and {Mastroserio}, Guglielmo and {Matt}, Giorgio and {Mori}, Kaya and {Petrucci}, Pierre-Olivier and {Podgorn{\'y}}, Jakub and {Poutanen}, Juri and {Steiner}, James F. and {Svoboda}, Ji{\v{r}}{\'\i} and {Taverna}, Roberto and {Tombesi}, Francesco and {Ravi}, Swati and {Rodriguez}, J{\'e}r{\^o}me and {Russell}, Thomas D. and {Veledina}, Alexandra and {Zhang}, Shuo},
        title = "{Exploring MAXI J1744-294: IXPE insights into a Newly Discovered X-ray Transient}",
      journal = {arXiv e-prints},
         year = 2025,
        month = jun,
          eid = {arXiv:2506.17050},
        pages = {arXiv:2506.17050},
          doi = {10.48550/arXiv.2506.17050},
archivePrefix = {arXiv},
       eprint = {2506.17050},
 primaryClass = {astro-ph.HE},
       adsurl = {https://ui.adsabs.harvard.edu/abs/2025arXiv250617050M}
}

@article{astropy:2013,
Adsurl = {http://adsabs.harvard.edu/abs/2013A%26A...558A..33A},
Archiveprefix = {arXiv},
Author = {{Astropy Collaboration} and {Robitaille}, T.~P. and {Tollerud}, E.~J. and {Greenfield}, P. and {Droettboom}, M. and {Bray}, E. and {Aldcroft}, T. and {Davis}, M. and {Ginsburg}, A. and {Price-Whelan}, A.~M. and {Kerzendorf}, W.~E. and {Conley}, A. and {Crighton}, N. and {Barbary}, K. and {Muna}, D. and {Ferguson}, H. and {Grollier}, F. and {Parikh}, M.~M. and {Nair}, P.~H. and {Unther}, H.~M. and {Deil}, C. and {Woillez}, J. and {Conseil}, S. and {Kramer}, R. and {Turner}, J.~E.~H. and {Singer}, L. and {Fox}, R. and {Weaver}, B.~A. and {Zabalza}, V. and {Edwards}, Z.~I. and {Azalee Bostroem}, K. and {Burke}, D.~J. and {Casey}, A.~R. and {Crawford}, S.~M. and {Dencheva}, N. and {Ely}, J. and {Jenness}, T. and {Labrie}, K. and {Lim}, P.~L. and {Pierfederici}, F. and {Pontzen}, A. and {Ptak}, A. and {Refsdal}, B. and {Servillat}, M. and {Streicher}, O.},
Doi = {10.1051/0004-6361/201322068},
Eid = {A33},
Eprint = {1307.6212},
Journal = {\aap},
Month = oct,
Pages = {A33},
Primaryclass = {astro-ph.IM},
Title = {{Astropy: A community Python package for astronomy}},
Volume = 558,
Year = 2013}

@ARTICLE{astropy:2018,
       author = {{Astropy Collaboration} and {Price-Whelan}, A.~M. and
         {Sip{\H{o}}cz}, B.~M. and {G{\"u}nther}, H.~M. and {Lim}, P.~L. and
         {Crawford}, S.~M. and {Conseil}, S. and {Shupe}, D.~L. and
         {Craig}, M.~W. and {Dencheva}, N. and {Ginsburg}, A. and {Vand
        erPlas}, J.~T. and {Bradley}, L.~D. and {P{\'e}rez-Su{\'a}rez}, D. and
         {de Val-Borro}, M. and {Aldcroft}, T.~L. and {Cruz}, K.~L. and
         {Robitaille}, T.~P. and {Tollerud}, E.~J. and {Ardelean}, C. and
         {Babej}, T. and {Bach}, Y.~P. and {Bachetti}, M. and {Bakanov}, A.~V. and
         {Bamford}, S.~P. and {Barentsen}, G. and {Barmby}, P. and
         {Baumbach}, A. and {Berry}, K.~L. and {Biscani}, F. and {Boquien}, M. and
         {Bostroem}, K.~A. and {Bouma}, L.~G. and {Brammer}, G.~B. and
         {Bray}, E.~M. and {Breytenbach}, H. and {Buddelmeijer}, H. and
         {Burke}, D.~J. and {Calderone}, G. and {Cano Rodr{\'\i}guez}, J.~L. and
         {Cara}, M. and {Cardoso}, J.~V.~M. and {Cheedella}, S. and {Copin}, Y. and
         {Corrales}, L. and {Crichton}, D. and {D'Avella}, D. and {Deil}, C. and
         {Depagne}, {\'E}. and {Dietrich}, J.~P. and {Donath}, A. and
         {Droettboom}, M. and {Earl}, N. and {Erben}, T. and {Fabbro}, S. and
         {Ferreira}, L.~A. and {Finethy}, T. and {Fox}, R.~T. and
         {Garrison}, L.~H. and {Gibbons}, S.~L.~J. and {Goldstein}, D.~A. and
         {Gommers}, R. and {Greco}, J.~P. and {Greenfield}, P. and
         {Groener}, A.~M. and {Grollier}, F. and {Hagen}, A. and {Hirst}, P. and
         {Homeier}, D. and {Horton}, A.~J. and {Hosseinzadeh}, G. and {Hu}, L. and
         {Hunkeler}, J.~S. and {Ivezi{\'c}}, {\v{Z}}. and {Jain}, A. and
         {Jenness}, T. and {Kanarek}, G. and {Kendrew}, S. and {Kern}, N.~S. and
         {Kerzendorf}, W.~E. and {Khvalko}, A. and {King}, J. and {Kirkby}, D. and
         {Kulkarni}, A.~M. and {Kumar}, A. and {Lee}, A. and {Lenz}, D. and
         {Littlefair}, S.~P. and {Ma}, Z. and {Macleod}, D.~M. and
         {Mastropietro}, M. and {McCully}, C. and {Montagnac}, S. and
         {Morris}, B.~M. and {Mueller}, M. and {Mumford}, S.~J. and {Muna}, D. and
         {Murphy}, N.~A. and {Nelson}, S. and {Nguyen}, G.~H. and
         {Ninan}, J.~P. and {N{\"o}the}, M. and {Ogaz}, S. and {Oh}, S. and
         {Parejko}, J.~K. and {Parley}, N. and {Pascual}, S. and {Patil}, R. and
         {Patil}, A.~A. and {Plunkett}, A.~L. and {Prochaska}, J.~X. and
         {Rastogi}, T. and {Reddy Janga}, V. and {Sabater}, J. and
         {Sakurikar}, P. and {Seifert}, M. and {Sherbert}, L.~E. and
         {Sherwood-Taylor}, H. and {Shih}, A.~Y. and {Sick}, J. and
         {Silbiger}, M.~T. and {Singanamalla}, S. and {Singer}, L.~P. and
         {Sladen}, P.~H. and {Sooley}, K.~A. and {Sornarajah}, S. and
         {Streicher}, O. and {Teuben}, P. and {Thomas}, S.~W. and
         {Tremblay}, G.~R. and {Turner}, J.~E.~H. and {Terr{\'o}n}, V. and
         {van Kerkwijk}, M.~H. and {de la Vega}, A. and {Watkins}, L.~L. and
         {Weaver}, B.~A. and {Whitmore}, J.~B. and {Woillez}, J. and
         {Zabalza}, V. and {Astropy Contributors}},
        title = "{The Astropy Project: Building an Open-science Project and Status of the v2.0 Core Package}",
      journal = {\aj},
         year = 2018,
        month = sep,
       volume = {156},
       number = {3},
          eid = {123},
        pages = {123},
          doi = {10.3847/1538-3881/aabc4f},
archivePrefix = {arXiv},
       eprint = {1801.02634},
 primaryClass = {astro-ph.IM},
       adsurl = {https://ui.adsabs.harvard.edu/abs/2018AJ....156..123A}
}

@ARTICLE{astropy:2022,
       author = {{Astropy Collaboration} and {Price-Whelan}, Adrian M. and {Lim}, Pey Lian and {Earl}, Nicholas and {Starkman}, Nathaniel and {Bradley}, Larry and {Shupe}, David L. and {Patil}, Aarya A. and {Corrales}, Lia and {Brasseur}, C.~E. and {N{"o}the}, Maximilian and {Donath}, Axel and {Tollerud}, Erik and {Morris}, Brett M. and {Ginsburg}, Adam and {Vaher}, Eero and {Weaver}, Benjamin A. and {Tocknell}, James and {Jamieson}, William and {van Kerkwijk}, Marten H. and {Robitaille}, Thomas P. and {Merry}, Bruce and {Bachetti}, Matteo and {G{"u}nther}, H. Moritz and {Aldcroft}, Thomas L. and {Alvarado-Montes}, Jaime A. and {Archibald}, Anne M. and {B{'o}di}, Attila and {Bapat}, Shreyas and {Barentsen}, Geert and {Baz{'a}n}, Juanjo and {Biswas}, Manish and {Boquien}, M{'e}d{'e}ric and {Burke}, D.~J. and {Cara}, Daria and {Cara}, Mihai and {Conroy}, Kyle E. and {Conseil}, Simon and {Craig}, Matthew W. and {Cross}, Robert M. and {Cruz}, Kelle L. and {D'Eugenio}, Francesco and {Dencheva}, Nadia and {Devillepoix}, Hadrien A.~R. and {Dietrich}, J{"o}rg P. and {Eigenbrot}, Arthur Davis and {Erben}, Thomas and {Ferreira}, Leonardo and {Foreman-Mackey}, Daniel and {Fox}, Ryan and {Freij}, Nabil and {Garg}, Suyog and {Geda}, Robel and {Glattly}, Lauren and {Gondhalekar}, Yash and {Gordon}, Karl D. and {Grant}, David and {Greenfield}, Perry and {Groener}, Austen M. and {Guest}, Steve and {Gurovich}, Sebastian and {Handberg}, Rasmus and {Hart}, Akeem and {Hatfield-Dodds}, Zac and {Homeier}, Derek and {Hosseinzadeh}, Griffin and {Jenness}, Tim and {Jones}, Craig K. and {Joseph}, Prajwel and {Kalmbach}, J. Bryce and {Karamehmetoglu}, Emir and {Ka{l}uszy{'n}ski}, Miko{l}aj and {Kelley}, Michael S.~P. and {Kern}, Nicholas and {Kerzendorf}, Wolfgang E. and {Koch}, Eric W. and {Kulumani}, Shankar and {Lee}, Antony and {Ly}, Chun and {Ma}, Zhiyuan and {MacBride}, Conor and {Maljaars}, Jakob M. and {Muna}, Demitri and {Murphy}, N.~A. and {Norman}, Henrik and {O'Steen}, Richard and {Oman}, Kyle A. and {Pacifici}, Camilla and {Pascual}, Sergio and {Pascual-Granado}, J. and {Patil}, Rohit R. and {Perren}, Gabriel I. and {Pickering}, Timothy E. and {Rastogi}, Tanuj and {Roulston}, Benjamin R. and {Ryan}, Daniel F. and {Rykoff}, Eli S. and {Sabater}, Jose and {Sakurikar}, Parikshit and {Salgado}, Jes{'u}s and {Sanghi}, Aniket and {Saunders}, Nicholas and {Savchenko}, Volodymyr and {Schwardt}, Ludwig and {Seifert-Eckert}, Michael and {Shih}, Albert Y. and {Jain}, Anany Shrey and {Shukla}, Gyanendra and {Sick}, Jonathan and {Simpson}, Chris and {Singanamalla}, Sudheesh and {Singer}, Leo P. and {Singhal}, Jaladh and {Sinha}, Manodeep and {Sip{H{o}}cz}, Brigitta M. and {Spitler}, Lee R. and {Stansby}, David and {Streicher}, Ole and {{{S}}umak}, Jani and {Swinbank}, John D. and {Taranu}, Dan S. and {Tewary}, Nikita and {Tremblay}, Grant R. and {Val-Borro}, Miguel de and {Van Kooten}, Samuel J. and {Vasovi{'c}}, Zlatan and {Verma}, Shresth and {de Miranda Cardoso}, Jos{'e} Vin{'i}cius and {Williams}, Peter K.~G. and {Wilson}, Tom J. and {Winkel}, Benjamin and {Wood-Vasey}, W.~M. and {Xue}, Rui and {Yoachim}, Peter and {Zhang}, Chen and {Zonca}, Andrea and {Astropy Project Contributors}},
        title = "{The Astropy Project: Sustaining and Growing a Community-oriented Open-source Project and the Latest Major Release (v5.0) of the Core Package}",
      journal = {\apj},
         year = 2022,
        month = aug,
       volume = {935},
       number = {2},
          eid = {167},
        pages = {167},
          doi = {10.3847/1538-4357/ac7c74},
archivePrefix = {arXiv},
       eprint = {2206.14220},
 primaryClass = {astro-ph.IM},
       adsurl = {https://ui.adsabs.harvard.edu/abs/2022ApJ...935..167A}
}

@ARTICLE{Brentjens2005,
       author = {{Brentjens}, M.~A. and {de Bruyn}, A.~G.},
        title = "{Faraday rotation measure synthesis}",
      journal = {\aap},
         year = 2005,
        month = oct,
       volume = {441},
       number = {3},
        pages = {1217-1228},
          doi = {10.1051/0004-6361:20052990},
archivePrefix = {arXiv},
       eprint = {astro-ph/0507349},
 primaryClass = {astro-ph},
       adsurl = {https://ui.adsabs.harvard.edu/abs/2005A&A...441.1217B}
}

@software{RMTools2020,
       author = {{Purcell}, C.~R. and {Van Eck}, C.~L. and {West}, J. and {Sun}, X.~H. and {Gaensler}, B.~M.},
        title = "{RM-Tools: Rotation measure (RM) synthesis and Stokes QU-fitting}",
 howpublished = {Astrophysics Source Code Library, record ascl:2005.003},
         year = 2020,
        month = may,
          eid = {ascl:2005.003},
archivePrefix = {ascl},
       eprint = {2005.003},
       adsurl = {https://ui.adsabs.harvard.edu/abs/2020ascl.soft05003P}
}

@software{dynesty,
  author       = {Sergey Koposov and
                  Josh Speagle and
                  Kyle Barbary and
                  Gregory Ashton and
                  Johannes Buchner and
                  Carl Scheffler and
                  Ben Cook and
                  Colm Talbot and
                  James Guillochon and
                  Patricio Cubillos and
                  Andrés Asensio Ramos and
                  Ben Johnson and
                  Dustin Lang and
                  Ilya and
                  Matthieu Dartiailh and
                  Alex Nitz and
                  Andrew McCluskey and
                  Anne Archibald and
                  Christoph Deil and
                  Dan Foreman-Mackey and
                  Danny Goldstein and
                  Erik Tollerud and
                  Joel Leja and
                  Matthew Kirk and
                  Matt Pitkin and
                  Patrick Sheehan and
                  Phillip Cargile and
                  ruskin23 and
                  Ruth Angus and
                  Tansu Daylan},
  title        = {joshspeagle/dynesty: v1.2.1},
  month        = apr,
  year         = 2022,
  publisher    = {Zenodo},
  version      = {v1.2.1},
  doi          = {10.5281/zenodo.6414759},
  url          = {https://doi.org/10.5281/zenodo.6414759},
}

@ARTICLE{Desvignes2018,
       author = {{Desvignes}, G. and {Eatough}, R.~P. and {Pen}, U.~L. and {Lee}, K.~J. and {Mao}, S.~A. and {Karuppusamy}, R. and {Schnitzeler}, D.~H.~F.~M. and {Falcke}, H. and {Kramer}, M. and {Wucknitz}, O. and {Spitler}, L.~G. and {Torne}, P. and {Liu}, K. and {Bower}, G.~C. and {Cognard}, I. and {Lyne}, A.~G. and {Stappers}, B.~W.},
        title = "{Large Magneto-ionic Variations toward the Galactic Center Magnetar, PSR J1745-2900}",
      journal = {\apjl},
         year = 2018,
        month = jan,
       volume = {852},
       number = {1},
          eid = {L12},
        pages = {L12},
          doi = {10.3847/2041-8213/aaa2f8},
archivePrefix = {arXiv},
       eprint = {1711.10323},
 primaryClass = {astro-ph.HE},
       adsurl = {https://ui.adsabs.harvard.edu/abs/2018ApJ...852L..12D}
}

@article{Dickey2019,
doi = {10.3847/1538-4357/aaf85f},
url = {https://doi.org/10.3847/1538-4357/aaf85f},
year = {2019},
month = {jan},
publisher = {The American Astronomical Society},
volume = {871},
number = {1},
pages = {106},
author = {Dickey, John M. and Landecker, T. L. and Thomson, Alec J. M. and Wolleben, M. and Sun, X. and Carretti, E. and Douglas, K. and Fletcher, A. and Gaensler, B. M. and Gray, A. and Haverkorn, M. and Hill, A. S. and Mao, S. A. and McClure-Griffiths, N. M.},
title = {The Galactic Magneto-ionic Medium Survey: Moments of the Faraday Spectra},
journal = {The Astrophysical Journal}
}

@ARTICLE{OSullivan2012,
       author = {{O'Sullivan}, S.~P. and {Brown}, S. and {Robishaw}, T. and {Schnitzeler}, D.~H.~F.~M. and {McClure-Griffiths}, N.~M. and {Feain}, I.~J. and {Taylor}, A.~R. and {Gaensler}, B.~M. and {Landecker}, T.~L. and {Harvey-Smith}, L. and {Carretti}, E.},
        title = "{Complex Faraday depth structure of active galactic nuclei as revealed by broad-band radio polarimetry}",
      journal = {\mnras},
         year = 2012,
        month = apr,
       volume = {421},
       number = {4},
        pages = {3300-3315},
          doi = {10.1111/j.1365-2966.2012.20554.x},
archivePrefix = {arXiv},
       eprint = {1201.3161},
 primaryClass = {astro-ph.CO},
       adsurl = {https://ui.adsabs.harvard.edu/abs/2012MNRAS.421.3300O}
}

@ARTICLE{Zdziarski2022,
       author = {{Zdziarski}, Andrzej A. and {Tetarenko}, Alexandra J. and {Sikora}, Marek},
        title = "{Jet Parameters in the Black Hole X-Ray Binary MAXI J1820+070}",
      journal = {\apj},
         year = 2022,
        month = feb,
       volume = {925},
       number = {2},
          eid = {189},
        pages = {189},
          doi = {10.3847/1538-4357/ac38a9},
archivePrefix = {arXiv},
       eprint = {2108.10929},
 primaryClass = {astro-ph.HE},
       adsurl = {https://ui.adsabs.harvard.edu/abs/2022ApJ...925..189Z}
}

@ARTICLE{Hughes2023,
       author = {{Hughes}, A.~K. and {Sivakoff}, G.~R. and {Macpherson}, C.~E. and {Miller-Jones}, J.~C.~A. and {Tetarenko}, A.~J. and {Altamirano}, D. and {Anderson}, G.~E. and {Belloni}, T.~M. and {Heinz}, S. and {Jonker}, P.~G. and {K{\"o}rding}, E.~G. and {Maitra}, D. and {Markoff}, S.~B. and {Migliari}, S. and {Mooley}, K.~P. and {Rupen}, M.~P. and {Russell}, D.~M. and {Russell}, T.~D. and {Sarazin}, C.~L. and {Soria}, R. and {Tudose}, V.},
        title = "{Short time-scale evolution of the polarized radio jet during V404 Cygni's 2015 outburst}",
      journal = {\mnras},
         year = 2023,
        month = may,
       volume = {521},
       number = {1},
        pages = {185-207},
          doi = {10.1093/mnras/stad396},
archivePrefix = {arXiv},
       eprint = {2301.13281},
 primaryClass = {astro-ph.HE},
       adsurl = {https://ui.adsabs.harvard.edu/abs/2023MNRAS.521..185H}
}

@BOOK{RybickiLightman1986,
       author = {{Rybicki}, George B. and {Lightman}, Alan P.},
        title = "{Radiative Processes in Astrophysics}",
         year = 1986,
       adsurl = {https://ui.adsabs.harvard.edu/abs/1986rpa..book.....R}
}

@BOOK{ERA2016,
       author = {{Condon}, James J. and {Ransom}, Scott M.},
        title = "{Essential Radio Astronomy}",
         year = 2016,
       adsurl = {https://ui.adsabs.harvard.edu/abs/2016era..book.....C}
}

@ARTICLE{MillerJones2019,
       author = {{Miller-Jones}, James C.~A. and {Tetarenko}, Alexandra J. and {Sivakoff}, Gregory R. and {Middleton}, Matthew J. and {Altamirano}, Diego and {Anderson}, Gemma E. and {Belloni}, Tomaso M. and {Fender}, Rob P. and {Jonker}, Peter G. and {K{\"o}rding}, Elmar G. and {Krimm}, Hans A. and {Maitra}, Dipankar and {Markoff}, Sera and {Migliari}, Simone and {Mooley}, Kunal P. and {Rupen}, Michael P. and {Russell}, David M. and {Russell}, Thomas D. and {Sarazin}, Craig L. and {Soria}, Roberto and {Tudose}, Valeriu},
        title = "{A rapidly changing jet orientation in the stellar-mass black-hole system V404 Cygni}",
      journal = {\nat},
         year = 2019,
        month = apr,
       volume = {569},
       number = {7756},
        pages = {374-377},
          doi = {10.1038/s41586-019-1152-0},
archivePrefix = {arXiv},
       eprint = {1906.05400},
 primaryClass = {astro-ph.HE},
       adsurl = {https://ui.adsabs.harvard.edu/abs/2019Natur.569..374M}
}

@ARTICLE{GRAVITY2019,
       author = {{GRAVITY Collaboration} and {Abuter}, R. and {Amorim}, A. and {Baub{\"o}ck}, M. and {Berger}, J.~P. and {Bonnet}, H. and {Brandner}, W. and {Cl{\'e}net}, Y. and {Coud{\'e} Du Foresto}, V. and {de Zeeuw}, P.~T. and {Dexter}, J. and {Duvert}, G. and {Eckart}, A. and {Eisenhauer}, F. and {F{\"o}rster Schreiber}, N.~M. and {Garcia}, P. and {Gao}, F. and {Gendron}, E. and {Genzel}, R. and {Gerhard}, O. and {Gillessen}, S. and {Habibi}, M. and {Haubois}, X. and {Henning}, T. and {Hippler}, S. and {Horrobin}, M. and {Jim{\'e}nez-Rosales}, A. and {Jocou}, L. and {Kervella}, P. and {Lacour}, S. and {Lapeyr{\`e}re}, V. and {Le Bouquin}, J.-B. and {L{\'e}na}, P. and {Ott}, T. and {Paumard}, T. and {Perraut}, K. and {Perrin}, G. and {Pfuhl}, O. and {Rabien}, S. and {Rodriguez Coira}, G. and {Rousset}, G. and {Scheithauer}, S. and {Sternberg}, A. and {Straub}, O. and {Straubmeier}, C. and {Sturm}, E. and {Tacconi}, L.~J. and {Vincent}, F. and {von Fellenberg}, S. and {Waisberg}, I. and {Widmann}, F. and {Wieprecht}, E. and {Wiezorrek}, E. and {Woillez}, J. and {Yazici}, S.},
        title = "{A geometric distance measurement to the Galactic center black hole with 0.3\% uncertainty}",
      journal = {\aap},
         year = 2019,
        month = may,
       volume = {625},
          eid = {L10},
        pages = {L10},
          doi = {10.1051/0004-6361/201935656},
archivePrefix = {arXiv},
       eprint = {1904.05721},
 primaryClass = {astro-ph.GA},
       adsurl = {https://ui.adsabs.harvard.edu/abs/2019A&A...625L..10G}
}

@ARTICLE{vonFellenberg2025,
       author = {{von Fellenberg}, Sebastiano D. and {Roychowdhury}, Tamojeet and {Michail}, Joseph M. and {Sumners}, Zach and {Sanger-Johnson}, Grace and {Fazio}, Giovanni G. and {Haggard}, Daryl and {Hora}, Joseph L. and {Philippov}, Alexander and {Ripperda}, Bart and {Smith}, Howard A. and {Willner}, S.~P. and {Witzel}, Gunther and {Zhang}, Shuo and {Becklin}, Eric E. and {Bower}, Geoffrey C. and {Chandra}, Sunil and {Do}, Tuan and {Garcia Marin}, Macarena and {Gurwell}, Mark A. and {Ford}, Nicole M. and {Hada}, Kazuhiro and {Markoff}, Sera and {Morris}, Mark R. and {Neilsen}, Joey and {Sabha}, Nadeen B. and {Seefeldt-Gail}, Braden},
        title = "{First Mid-infrared Detection and Modeling of a Flare from Sgr A*}",
      journal = {\apjl},
         year = 2025,
        month = jan,
       volume = {979},
       number = {1},
          eid = {L20},
        pages = {L20},
          doi = {10.3847/2041-8213/ada3d2},
archivePrefix = {arXiv},
       eprint = {2501.07415},
 primaryClass = {astro-ph.HE},
       adsurl = {https://ui.adsabs.harvard.edu/abs/2025ApJ...979L..20V}
}

@ARTICLE{Michail2025,
       author = {{Michail}, Joseph M. and {von Fellenberg}, Sebastiano D. and {Keating}, Garrett K. and {Rao}, Ramprasad and {Roychowdhury}, Tamojeet and {Willner}, S.~P. and {Ford}, Nicole M. and {Haggard}, Daryl and {Markoff}, Sera and {Philippov}, Alexander and {Ripperda}, Bart and {Sanchez-Maes}, Sophia and {Sumners}, Zach and {Witzel}, Gunther and {Balakrishnan}, Mayura and {Chandra}, Sunil and {Hada}, Kazuhiro and {Garcia Marin}, Macarena and {Gurwell}, Mark A. and {Fazio}, Giovanni G. and {Hora}, Joseph L. and {Seefeldt-Gail}, Braden and {Smith}, Howard A.},
        title = "{First Mid-infrared Detection and Modeling of a Flare from Sgr A*. II. Mid-infrared Spectral Energy Distribution and Millimeter Polarimetry}",
      journal = {\apj},
         year = 2026,
        month = feb,
       volume = {997},
       number = {2},
          eid = {282},
        pages = {282},
          doi = {10.3847/1538-4357/ae25ef},
archivePrefix = {arXiv},
       eprint = {2511.14836},
 primaryClass = {astro-ph.HE},
       adsurl = {https://ui.adsabs.harvard.edu/abs/2026ApJ...997..282M}
}

@ARTICLE{Roychowdhury2025,
       author = {{Roychowdhury}, Tamojeet and {von Fellenberg}, Sebastiano D. and {Michail}, Joseph M. and {Willner}, S.~P. and {Ford}, Nicole M. and {Sumners}, Zach and {Sanchez-Maes}, Sophia and {Do}, Tuan and {Garcia Marin}, Macarena and {Markoff}, Sera and {Fazio}, Giovanni G. and {Haggard}, Daryl and {Hora}, Joseph L. and {Ripperda}, Bart and {Sabha}, Nadeen B. and {Smith}, Howard A. and {Witzel}, Gunther},
        title = "{Photometric Constraints on Intermediate-mass Black Holes in the Galactic Centre}",
      journal = {\pasp},
         year = 2025,
        month = nov,
       volume = {137},
       number = {11},
          eid = {114102},
        pages = {114102},
          doi = {10.1088/1538-3873/ae16d6},
archivePrefix = {arXiv},
       eprint = {2511.14856},
 primaryClass = {astro-ph.GA},
       adsurl = {https://ui.adsabs.harvard.edu/abs/2025PASP..137k4102R}
}

@ARTICLE{Sault1996,
       author = {{Sault}, R.~J. and {Hamaker}, J.~P. and {Bregman}, J.~D.},
        title = "{Understanding radio polarimetry. II. Instrumental calibration of an interferometer array.}",
      journal = {\aaps},
         year = 1996,
        month = may,
       volume = {117},
        pages = {149-159},
       adsurl = {https://ui.adsabs.harvard.edu/abs/1996A&AS..117..149S}
}

@ARTICLE{Bower1999_lo,
       author = {{Bower}, Geoffrey C. and {Backer}, Donald C. and {Zhao}, Jun-Hui and {Goss}, Miller and {Falcke}, Heino},
        title = "{The Linear Polarization of Sagittarius A*. I. VLA Spectropolarimetry at 4.8 and 8.4 GHZ}",
      journal = {\apj},
         year = 1999,
        month = aug,
       volume = {521},
       number = {2},
        pages = {582-586},
          doi = {10.1086/307592},
archivePrefix = {arXiv},
       eprint = {astro-ph/9904091},
 primaryClass = {astro-ph},
       adsurl = {https://ui.adsabs.harvard.edu/abs/1999ApJ...521..582B}
}

@ARTICLE{Bower2018,
       author = {{Bower}, Geoffrey C. and {Broderick}, Avery and {Dexter}, Jason and {Doeleman}, Shepherd and {Falcke}, Heino and {Fish}, Vincent and {Johnson}, Michael D. and {Marrone}, Daniel P. and {Moran}, James M. and {Moscibrodzka}, Monika and {Peck}, Alison and {Plambeck}, Richard L. and {Rao}, Ramprasad},
        title = "{ALMA Polarimetry of Sgr A*: Probing the Accretion Flow from the Event Horizon to the Bondi Radius}",
      journal = {\apj},
         year = 2018,
        month = dec,
       volume = {868},
       number = {2},
          eid = {101},
        pages = {101},
          doi = {10.3847/1538-4357/aae983},
archivePrefix = {arXiv},
       eprint = {1810.07317},
 primaryClass = {astro-ph.HE},
       adsurl = {https://ui.adsabs.harvard.edu/abs/2018ApJ...868..101B}
}

@ARTICLE{Liu2016,
       author = {{Liu}, Hauyu Baobab and {Wright}, Melvyn C.~H. and {Zhao}, Jun-Hui and {Brinkerink}, Christiaan D. and {Ho}, Paul T.~P. and {Mills}, Elisabeth A.~C. and {Mart{\'\i}n}, Sergio and {Falcke}, Heino and {Matsushita}, Satoki and {Mart{\'\i}-Vidal}, Ivan},
        title = "{Linearly polarized millimeter and submillimeter continuum emission of Sgr A* constrained by ALMA}",
      journal = {\aap},
         year = 2016,
        month = sep,
       volume = {593},
          eid = {A107},
        pages = {A107},
          doi = {10.1051/0004-6361/201628731},
archivePrefix = {arXiv},
       eprint = {1605.05544},
 primaryClass = {astro-ph.HE},
       adsurl = {https://ui.adsabs.harvard.edu/abs/2016A&A...593A.107L}
}

@ARTICLE{YusefZadeh2015,
       author = {{Yusef-Zadeh}, F. and {Diesing}, R. and {Wardle}, M. and {Sjouwerman}, L.~O. and {Royster}, M. and {Cotton}, W.~D. and {Roberts}, D. and {Heinke}, C.},
        title = "{Radio Continuum Emission from the Magnetar SGR J1745-2900: Interaction with Gas Orbiting Sgr A*}",
      journal = {\apjl},
         year = 2015,
        month = oct,
       volume = {811},
       number = {2},
          eid = {L35},
        pages = {L35},
          doi = {10.1088/2041-8205/811/2/L35},
archivePrefix = {arXiv},
       eprint = {1509.03337},
 primaryClass = {astro-ph.HE},
       adsurl = {https://ui.adsabs.harvard.edu/abs/2015ApJ...811L..35Y}
}

@ARTICLE{Marrone2006,
       author = {{Marrone}, Daniel P. and {Moran}, James M. and {Zhao}, Jun-Hui and {Rao}, Ramprasad},
        title = "{Interferometric Measurements of Variable 340 GHz Linear Polarization in Sagittarius A*}",
      journal = {\apj},
         year = 2006,
        month = mar,
       volume = {640},
       number = {1},
        pages = {308-318},
          doi = {10.1086/500106},
archivePrefix = {arXiv},
       eprint = {astro-ph/0511653},
 primaryClass = {astro-ph},
       adsurl = {https://ui.adsabs.harvard.edu/abs/2006ApJ...640..308M}
}

@ARTICLE{Marrone2007,
       author = {{Marrone}, Daniel P. and {Moran}, James M. and {Zhao}, Jun-Hui and {Rao}, Ramprasad},
        title = "{An Unambiguous Detection of Faraday Rotation in Sagittarius A*}",
      journal = {\apjl},
         year = 2007,
        month = jan,
       volume = {654},
       number = {1},
        pages = {L57-L60},
          doi = {10.1086/510850},
archivePrefix = {arXiv},
       eprint = {astro-ph/0611791},
 primaryClass = {astro-ph},
       adsurl = {https://ui.adsabs.harvard.edu/abs/2007ApJ...654L..57M}
}

@ARTICLE{Bower2003,
       author = {{Bower}, Geoffrey C. and {Wright}, Melvyn C.~H. and {Falcke}, Heino and {Backer}, Donald C.},
        title = "{Interferometric Detection of Linear Polarization from Sagittarius A* at 230 GHz}",
      journal = {\apj},
         year = 2003,
        month = may,
       volume = {588},
       number = {1},
        pages = {331-337},
          doi = {10.1086/373989},
archivePrefix = {arXiv},
       eprint = {astro-ph/0302227},
 primaryClass = {astro-ph},
       adsurl = {https://ui.adsabs.harvard.edu/abs/2003ApJ...588..331B}
}

@ARTICLE{Abbate2023,
       author = {{Abbate}, F. and {Noutsos}, A. and {Desvignes}, G. and {Wharton}, R.~S. and {Torne}, P. and {Kramer}, M. and {Eatough}, R.~P. and {Karuppusamy}, R. and {Liu}, K. and {Shao}, L. and {Wongphechauxsorn}, J.},
        title = "{Rotation measure variations in Galactic Centre pulsars}",
      journal = {\mnras},
         year = 2023,
        month = sep,
       volume = {524},
       number = {2},
        pages = {2966-2977},
          doi = {10.1093/mnras/stad2047},
archivePrefix = {arXiv},
       eprint = {2307.03230},
 primaryClass = {astro-ph.HE},
       adsurl = {https://ui.adsabs.harvard.edu/abs/2023MNRAS.524.2966A}
}

@ARTICLE{Roy2008,
       author = {{Roy}, S. and {Pramesh Rao}, A. and {Subrahmanyan}, R.},
        title = "{Magnetic field near the central region of the Galaxy: rotation measure of extragalactic sources}",
      journal = {\aap},
         year = 2008,
        month = feb,
       volume = {478},
       number = {2},
        pages = {435-442},
          doi = {10.1051/0004-6361:20066470},
archivePrefix = {arXiv},
       eprint = {0712.0269},
 primaryClass = {astro-ph},
       adsurl = {https://ui.adsabs.harvard.edu/abs/2008A&A...478..435R}
}

@ARTICLE{Sicheneder2017,
       author = {{Sicheneder}, Egid and {Dexter}, Jason},
        title = "{A single H II region model of the strong interstellar scattering towards Sgr A*}",
      journal = {\mnras},
         year = 2017,
        month = may,
       volume = {467},
       number = {3},
        pages = {3642-3647},
          doi = {10.1093/mnras/stx103},
archivePrefix = {arXiv},
       eprint = {1612.04819},
 primaryClass = {astro-ph.GA},
       adsurl = {https://ui.adsabs.harvard.edu/abs/2017MNRAS.467.3642S}
}

@ARTICLE{Mori2019,
       author = {{Mori}, Kaya and {Hailey}, Charles J. and {Mandel}, Shifra and {Schutt}, Theo and {Bachetti}, Matteo and {Coerver}, Anna and {Baganoff}, Frederick K. and {Dykaar}, Hannah and {Grindlay}, Jonathan E. and {Haggard}, Daryl and {Heuer}, Keri and {Hong}, Jaesub and {Hord}, Benjamin J. and {Jin}, Chichuan and {Nynka}, Melania and {Ponti}, Gabriele and {Tomsick}, John A.},
        title = "{NuSTAR and Chandra Observations of New X-Ray Transients in the Central Parsec of the Galaxy}",
      journal = {\apj},
         year = 2019,
        month = nov,
       volume = {885},
       number = {2},
          eid = {142},
        pages = {142},
          doi = {10.3847/1538-4357/ab4b47},
archivePrefix = {arXiv},
       eprint = {1910.03459},
 primaryClass = {astro-ph.HE},
       adsurl = {https://ui.adsabs.harvard.edu/abs/2019ApJ...885..142M}
}

@ARTICLE{Degenaar2016b,
       author = {{Degenaar}, N. and {Reynolds}, M.~T. and {Wijnands}, R. and {Miller}, J.~M. and {Kennea}, J.~A. and {Ponti}, G. and {Haggard}, D. and {Gehrels}, N.},
        title = "{Continued Swift/XRT observations of the new Galactic center transients SWIFT J174540.2-290037 and SWIFT J174540.7-290015}",
      journal = {The Astronomer's Telegram},
         year = 2016,
        month = jun,
       volume = {9196},
        pages = {1},
       adsurl = {https://ui.adsabs.harvard.edu/abs/2016ATel.9196....1D}
}

@ARTICLE{Degenaar2016a,
       author = {{Degenaar}, N. and {Reynolds}, M.~T. and {Wijnands}, R. and {Miller}, J.~M. and {Kennea}, J.~A. and {Ponti}, G. and {Haggard}, D. and {Gehrels}, N.},
        title = "{Swift/XRT detection of another active X-ray transient close to Sgr A*}",
      journal = {The Astronomer's Telegram},
         year = 2016,
        month = jun,
       volume = {9109},
        pages = {1},
       adsurl = {https://ui.adsabs.harvard.edu/abs/2016ATel.9109....1D}
}

@ARTICLE{Lacy2020,
       author = {{Lacy}, M. and {Baum}, S.~A. and {Chandler}, C.~J. and {Chatterjee}, S. and {Clarke}, T.~E. and {Deustua}, S. and {English}, J. and {Farnes}, J. and {Gaensler}, B.~M. and {Gugliucci}, N. and {Hallinan}, G. and {Kent}, B.~R. and {Kimball}, A. and {Law}, C.~J. and {Lazio}, T.~J.~W. and {Marvil}, J. and {Mao}, S.~A. and {Medlin}, D. and {Mooley}, K. and {Murphy}, E.~J. and {Myers}, S. and {Osten}, R. and {Richards}, G.~T. and {Rosolowsky}, E. and {Rudnick}, L. and {Schinzel}, F. and {Sivakoff}, G.~R. and {Sjouwerman}, L.~O. and {Taylor}, R. and {White}, R.~L. and {Wrobel}, J. and {Andernach}, H. and {Beasley}, A.~J. and {Berger}, E. and {Bhatnager}, S. and {Birkinshaw}, M. and {Bower}, G.~C. and {Brandt}, W.~N. and {Brown}, S. and {Burke-Spolaor}, S. and {Butler}, B.~J. and {Comerford}, J. and {Demorest}, P.~B. and {Fu}, H. and {Giacintucci}, S. and {Golap}, K. and {G{\"u}th}, T. and {Hales}, C.~A. and {Hiriart}, R. and {Hodge}, J. and {Horesh}, A. and {Ivezi{\'c}}, {\v{Z}}. and {Jarvis}, M.~J. and {Kamble}, A. and {Kassim}, N. and {Liu}, X. and {Loinard}, L. and {Lyons}, D.~K. and {Masters}, J. and {Mezcua}, M. and {Moellenbrock}, G.~A. and {Mroczkowski}, T. and {Nyland}, K. and {O'Dea}, C.~P. and {O'Sullivan}, S.~P. and {Peters}, W.~M. and {Radford}, K. and {Rao}, U. and {Robnett}, J. and {Salcido}, J. and {Shen}, Y. and {Sobotka}, A. and {Witz}, S. and {Vaccari}, M. and {van Weeren}, R.~J. and {Vargas}, A. and {Williams}, P.~K.~G. and {Yoon}, I.},
        title = "{The Karl G. Jansky Very Large Array Sky Survey (VLASS). Science Case and Survey Design}",
      journal = {\pasp},
         year = 2020,
        month = mar,
       volume = {132},
       number = {1009},
          eid = {035001},
        pages = {035001},
          doi = {10.1088/1538-3873/ab63eb},
archivePrefix = {arXiv},
       eprint = {1907.01981},
 primaryClass = {astro-ph.IM},
       adsurl = {https://ui.adsabs.harvard.edu/abs/2020PASP..132c5001L}
}

@ARTICLE{Mandel2026_atel,
       author = {{Mandel}, S. and {Mori}, K. and {Hua}, Z. and {Reynolds}, M. and {Degenaar}, N. and {Marra}, L. and {Parra}, M. and {Draghis}, P.},
        title = "{Determination of MAXI J1744-294 as a repeat outburst of the 2016 transient Swift J174540.2-290037}",
      journal = {The Astronomer's Telegram},
         year = 2026,
        month = feb,
       volume = {17663},
        pages = {1},
       adsurl = {https://ui.adsabs.harvard.edu/abs/2026ATel17663....1M}
}

@BOOK{Longair2011,
       author = {{Longair}, Malcolm S.},
        title = "{High Energy Astrophysics}",
         year = 2011,
       adsurl = {https://ui.adsabs.harvard.edu/abs/2011hea..book.....L}
}

@ARTICLE{Brocksopp2013,
       author = {{Brocksopp}, C. and {Corbel}, S. and {Tzioumis}, A. and {Broderick}, J.~W. and {Rodriguez}, J. and {Yang}, J. and {Fender}, R.~P. and {Paragi}, Z.},
        title = "{XTE J1752-223 in outburst: a persistent radio jet, dramatic flaring, multiple ejections and linear polarization}",
      journal = {\mnras},
         year = 2013,
        month = jun,
       volume = {432},
       number = {2},
        pages = {931-943},
          doi = {10.1093/mnras/stt493},
archivePrefix = {arXiv},
       eprint = {1303.6702},
 primaryClass = {astro-ph.HE},
       adsurl = {https://ui.adsabs.harvard.edu/abs/2013MNRAS.432..931B}
}

@ARTICLE{Hannikainen2000,
       author = {{Hannikainen}, D.~C. and {Hunstead}, R.~W. and {Campbell-Wilson}, D. and {Wu}, K. and {McKay}, D.~J. and {Smits}, D.~P. and {Sault}, R.~J.},
        title = "{Radio Emission from GRO J1655-40 during the 1994 Jet Ejection Episodes}",
      journal = {\apj},
         year = 2000,
        month = sep,
       volume = {540},
       number = {1},
        pages = {521-534},
          doi = {10.1086/309294},
archivePrefix = {arXiv},
       eprint = {astro-ph/0003466},
 primaryClass = {astro-ph},
       adsurl = {https://ui.adsabs.harvard.edu/abs/2000ApJ...540..521H}
}

@ARTICLE{Blandford1979,
       author = {{Blandford}, R.~D. and {K{\"o}nigl}, A.},
        title = "{Relativistic jets as compact radio sources.}",
      journal = {\apj},
         year = 1979,
        month = aug,
       volume = {232},
        pages = {34-48},
          doi = {10.1086/157262},
       adsurl = {https://ui.adsabs.harvard.edu/abs/1979ApJ...232...34B}
}

@ARTICLE{Rodriguez2024,
       author = {{Rodriguez Cavero}, Nicole},
        title = "{First Year of Stellar-Mass Black Hole Observations with the Imaging X-ray Polarimetry Explorer}",
      journal = {arXiv e-prints},
         year = 2024,
        month = feb,
          eid = {arXiv:2402.10371},
        pages = {arXiv:2402.10371},
          doi = {10.48550/arXiv.2402.10371},
archivePrefix = {arXiv},
       eprint = {2402.10371},
 primaryClass = {astro-ph.HE},
       adsurl = {https://ui.adsabs.harvard.edu/abs/2024arXiv240210371R}
}
